\documentclass[11pt,a4paper,pdftex]{article}
\pdfoutput=1
\usepackage{jcappub}
\usepackage{color}

\newcommand{\be}{\begin{eqnarray}}
\newcommand{\ee}{\end{eqnarray}}
\newcommand{\rar}{\rightarrow}

\title{Distinguishing black holes and naked singularities with iron line spectroscopy}

\author[a]{Honghui~Liu,}
\author[a]{Menglei~Zhou,}
\author[a,b,1]{Cosimo~Bambi%
\note{Corresponding author}}

\affiliation[a]{Center for Field Theory and Particle Physics and Department of Physics,\\
Fudan University, 2005 Songhu Road, 200438 Shanghai, China}
\affiliation[b]{Theoretical Astrophysics, Eberhard-Karls Universit\"at T\"ubingen,\\ 
Auf der Morgenstelle 10, 72076 T\"ubingen, Germany}

\emailAdd{hhliu15@fudan.edu.cn}
\emailAdd{mlzhou13@fudan.edu.cn}
\emailAdd{bambi@fudan.edu.cn}

\abstract{It is commonly thought that the final product of gravitational collapse is a black hole. Nevertheless, theoretical studies have not yet provided a final answer to the question whether black holes are the only possible outcome or whether naked singularities are also allowed. Observational tests may thus represent a complementary approach. In the present paper, we consider the Janis-Newman-Winicour metric, which describes a rotating source with a surface-like naked singularity. We calculate iron line shapes in the reflection spectrum of a putative disk around a Janis-Newman-Winicour singularity and we compare our results with the iron line shapes expected in the spectrum of a Kerr black hole. While it is difficult to distinguish the two spacetimes from the iron line shape in general, it seems that Janis-Newman-Winicour singularities cannot mimic fast-rotating Kerr black holes observed at a low or moderate inclination angle. Our analysis thus suggests that available observations of specific sources can already constrain the possible existence of Janis-Newman-Winicour singularities in the Universe.}


\begin{document}

\maketitle


\section{Introduction}

An important unsolved problem in Einstein's gravity concerns the nature of the final product of gravitational collapse. At the end of the 1960s, Roger Penrose and Stephen Hawking proved, under quite general assumptions, that complete gravitational collapse inevitably produces a spacetime singularity~\cite{s1,s2}. Penrose also proposed the so-called {\it cosmic censorship conjecture}, according to which singularities produced in gravitational collapse must be hidden behind an event horizon and the final product of the collapse must be a black hole~\cite{ccc}. Nevertheless, today we know a few counterexamples in which naked singularities can be created from regular initial data~\cite{sr}. While the singularity itself may not exist because it may be an artifact of the breakdown of the classical theory, the spacetime metric around the source may still be described by the classical singular solution~\cite{super}.

Today we have a large number of astronomical observations pointing out the existence of dark and compact objects that are commonly interpreted as black holes~\cite{book,rev}. From stellar evolution arguments, we expect a population of $10^8-10^9$~stellar-mass black holes in a galaxy like the Milky Way~\cite{bh1,bh2}. Today we only know about 100~candidates. Observations also suggest the existence of supermassive black holes of millions or billions Solar masses at the center of every normal galaxy. All these objects are commonly interpreted as black holes because this is the simplest explanation. Stellar-mass black holes are too heavy to be neutron stars~\cite{lattimer}. Supermassive black holes are too heavy, compact, and old to be clusters of non-luminous bodies like neutron stars~\cite{maoz}. The non-detection of thermal radiation from the putative surface of these objects is also consistent with the idea that they are black holes with an event horizon~\cite{hor}. The gravitational wave signals detected by the LIGO experiment are consistent with gravitational waves emitted by black holes~\cite{ligoligo}, even if this is not enough to conclude that these sources have a horizon~\cite{gw}.

In the past 5-10~years, there have been significant efforts to study observational methods to probe the spacetime metric around astrophysical black hole candidates and test the actual nature of these objects using the electromagnetic radiation emitted by their accretion disk~\cite{t1,t2,t2b,t2c,t2d,t3,t4,t5,t5b,t6,t7,t7b,t8,t9,t10,r1,r2,r2b,r3}. These tests are sensitive to the motion of the particles in the accretion disk and to the propagation of the photons from the emission point in the disk to the detection point in the flat faraway region. The detected spectrum is affected by relativistic effects occurring in the strong gravity region (gravitational redshift, Doppler boosting, light bending) and can thus provide details of the metric around these objects.

In the present paper, we extend previous work in literature to study observational tests to distinguish black holes from naked singularities~\cite{x1,x2,x3,x4,x5}. We consider the rotating Janis-Newman-Winicour (JNW) metric~\cite{jnw1,jnw2,jnw3}, which describes a rotating source in Einstein's gravity minimally coupled to a real scalar field. The Schwarzschild metric is the unique vacuum solution to the 4-dimensional Einstein equations after we impose that the spacetime is spherically symmetric and asymptotically flat. If we relax the assumption of vacuum spacetime and we introduce a real scalar field, which is the simplest form of matter, we obtain the non-rotating JNW metric (if we introduce an electromagnetic field, we find instead the more famous Reissner-Nordstr\"om metric). The rotating JNW solution is thus one of the simplest generalizations of the Kerr metric and includes the Kerr solution as a special case (in the limit of vanishing scalar charge). Interestingly, the JNW spacetime does not posses any horizon and there is a surface-like naked singularity at a finite value of the radial coordinate, where curvature invariants diverge and the spacetime is geodetically incomplete.

We calculate iron line shapes expected in the reflection spectrum of an accretion disk in JNW spacetimes and we compare our results with the iron line shapes in the reflection spectrum of Kerr black holes\footnote{We note that iron line shapes expected in the reflection spectrum of accretion disks around Kerr naked singularities have been studied in Ref.~\cite{stuck}. As we will see later, the iron line shapes from JNW naked singularities are qualitatively different from those from Kerr naked singularities.}. Our simple analysis shows that iron line shapes in JNW spacetimes can in general reasonably well mimic those expected in Kerr spacetimes. However, they cannot look like those from accretion disks observed from a low or moderate inclination angle around fast-rotating black holes. Since we have X-ray data of astrophysical black hole candidates showing iron line shapes as expected from fast-rotating black holes observed at low or moderate inclination angles, our results suggest that current data can already rule out, or at least strongly constrain, the possibility that astrophysical black holes are actually JNW naked singularities. Note, however, that the iron line method has a number of caveats\footnote{Current models employ infinitesimally thin accretion disks with a sharp inner edge at the innermost stable circular orbit (ISCO), while real accretion disks have a finite thickness (which increases as the mass accretion rate increases), the inner edge may not be exactly at the ISCO, and radiation may be emitted even at smaller radii by plunging gas. The emissivity profile of the disk, which would depend on the unknown morphology of the Comptonized medium, is usually modeled by a power-law or a broken power-law, and is often one of the most controversial issues for the reliability of the technique. The presence of magnetic fields is ignored. A number of simplifications are also employed in the calculations of the reflection spectrum at the emission point (cold gas of constant density, fixed electron density in the disk, constant ionization over the disk, etc.). See \cite{book} and references therein for more details.} and that, even assuming to have the systematics completely under control, a conclusive answer would require a more detailed study, beyond the goal of the present paper, and to analyze real data with a more sophisticated reflection model.

The content of the paper is as follows. In Section~\ref{s-jnw}, we briefly review the JNW metric and its basic properties. In Section~\ref{s-iron}, we introduce the iron line method and we calculate a set of iron line shapes from a putative accretion disk around JNW naked singularities for different values of the spin parameter $a_*$, inclination angle of the disk $i$, and deformation parameter $\gamma$. In Section~\ref{s-sim}, we simulate some observations with NuSTAR and we check whether the analysis of the iron line can distinguish Kerr black holes from JNW naked singularities. Summary and conclusions are in Section~\ref{s-con}. Throughout the paper, we employ units in which $G_{\rm N} = c = 1$ and a metric with signature $(-+++)$.


\section{Janis-Newman-Winicour spacetime \label{s-jnw}}

\subsection{Metric}

The JNW spacetime is an exact solution of the Einstein equations in which matter is described by a real massless scalar field $\Phi$~\cite{jnw1,jnw2,jnw3}. The total action is
\be
S = \int d^4 x \, \sqrt{-g} \left[ R + g^{\mu\nu} 
\left(\partial_\mu \Phi\right) \left(\partial_\nu \Phi\right) \right] \, .
\ee
The field equations are
\be
R_{\mu\nu} = 8 \left(\partial_\mu \Phi\right) \left(\partial_\nu \Phi\right) \, , \quad
\Box \Phi = 0 \, .
\ee
The line element of the spacetime can be written as 
\be
ds^2 &=& \left( 1 - \frac{2 \tilde{M} r}{\Sigma} \right)^\gamma 
\left( dt - \omega d\phi \right)^2 
 - \Sigma \left( 1 - \frac{2 \tilde{M} r}{\Sigma} \right)^{1 - \gamma} 
\left( \frac{dr^2}{\Delta} + d\theta^2 + \sin^2\theta d\phi^2 \right) \nonumber\\
&& + 2 \omega \left(dt - \omega d\phi\right) d\phi \, ,
\ee
where
\be
\tilde{M} = \frac{M}{\gamma} \, , \quad
\omega = a \sin^2\theta \, , \quad
\Sigma = r^2 + a^2 \cos^2\theta \, , \quad
\Delta = r^2 - 2 \tilde{M} r + a^2 \, . 
\ee
The solution for the scalar field is
\be
\Phi = \frac{\sqrt{1 - \gamma^2}}{4} 
\ln \left( 1 - \frac{2 \tilde{M} r}{\Sigma} \right) \, .
\ee
The parameter $\gamma$ is related to the ADM mass\footnote{For the definition of ADM mass, see, for instance, \cite{adm1,adm2,adm3}.} $M$ and the scalar charge $q$ by the following relation
\be
\gamma = \frac{M}{\sqrt{M^2 + q^2}} \, .
\ee
$a=J/M$, where $J$ is the spin angular momentum of the source. $a_* = a/M = J/M^2$ is the dimensionless spin parameter.

For $q=0$, $\gamma = 1$, and we recover the Kerr black hole solution. In this case, the larger root of $\Delta = 0$ provides the radius of the event horizon
\be\label{eq-bh}
r_{\rm H} = M + \sqrt{M^2 - a^2} \, .
\ee
The boundary of the ergoregion (static limit) is given by the larger root of $g_{tt} = 0$ and turns out to be
\be
r_{\rm SL} = M + \sqrt{M^2 - a^2 \cos^2\theta} \, .
\ee

For $q\neq0$, $0 < \gamma < 1$. The scalar curvature is
\be
R = \frac{2 \left( \gamma^2 - 1 \right) \tilde{M}^2}{\Sigma^5} 
\left( 1 - \frac{2 \tilde{M} r}{\Sigma} \right)^{\gamma - 3}
 \left[ \Delta \left( r^2 - a^2 \cos^2\theta \right)^2 
+ \left( r a^2 \sin^22\theta \right)^2 \right] \, , 
\ee
and diverges on the surface
\be\label{eq-ns}
r_* = \tilde{M} + \sqrt{\tilde{M}^2 - a^2 \cos^2\theta} \, .
\ee
$r_*$ describes a surface-like singularity, where curvature invariants diverge and the spacetime is geodetically incomplete as well.

The JNW spacetime can be seen as a deformation of the Kerr spacetime in which $\gamma$ is the parameter regulating this deformation. Note, however, that the transition from a black hole $(\gamma = 1)$ to a naked singularity $(\gamma < 1)$ is not smooth. This fact, as we will see later, has important observational implications and -- potentially -- observational data can completely rule out the existence of JNW naked singularities, not just constrain $\gamma$. The key-feature is that the boundary of a Kerr black hole is given by the radius of its event horizon in Eq.~(\ref{eq-bh}). The boundary of a JNW naked singularity is given by the radius of its singular surface in Eq.~(\ref{eq-ns}). For very fast-rotating objects with $a_*$ very close to 1, the inner edge of an equatorial accretion disk is, respectively, at the radial coordinate $r \approx M$ and $r \approx 2M$ in the two spacetimes; that is, there is a gap between the cases $\gamma = 1$ and $\gamma \neq 1$. This happens because for $\gamma \neq 1$ the static limit becomes the ``surface'' of the object, not the former event horizon. Note also that a JNW naked singularity is completely different from a Kerr naked singularity. The former can be seen as an extended object, like a compact star or a black hole. The latter is an almost point-like source.

\subsection{Location of the ISCO radius}

In the next sections, we will consider geometrically thin accretion disks in the equatorial plane $(\theta = \pi/2)$ of JNW naked singularities. The inner edge of the disk will be set at the radius of the innermost stable circular orbit (ISCO), if it exists, or, otherwise, at the singular surface of the object $r_*$. Circular equatorial orbits and accretion disks in the JNW spacetime have been already studied in Ref.~\cite{diskk}.

The calculation of the ISCO radius is straightforward~\cite{book}. We write the Lagrangian of a test-particle as
\be
\mathcal{L} = \frac{1}{2} g_{\mu\nu} \dot{x}^\mu \dot{x}^\nu \, ,
\ee 
where $\dot{} = d/d\tau$ and $\tau$ is the particle proper time. Since the metric coefficients are independent of the coordinates $t$ and $\phi$, we have the conservation of the energy $E$ and of the axial component of the angular momentum $L_z$. From the Euler-Lagrange equations, we find
\be
g_{tt} \dot{t} + g_{t\phi} \dot{\phi} = - E \, , \quad
g_{t\phi} \dot{t} + g_{\phi\phi} \dot{\phi} = L_z \, .
\ee
From the conservation of the rest-mass, $g_{\mu\nu} \dot{x}^\mu \dot{x}^\nu = -1$, we can write
\be
g_{rr} \dot{r}^2 + g_{\theta\theta} \dot{\theta}^2 = V_{\rm eff} \, ,
\ee
where the effective potential $V_{\rm eff}$ is given by
\be
V_{\rm eff} = \frac{E^2 g_{\phi\phi} + 2 E L_z g_{t\phi} 
+ L_z^2 g_{tt}}{g_{t\phi}^2 - g_{tt} g_{\phi\phi}} - 1 \, .
\ee
Circular orbits in the equatorial plane are located at the zeros and the turning points of the effective potential: $\dot{r}=\dot{\theta}=0$ implies $V_{\rm eff} = 0$, $\ddot{r}=0$ implies $\partial_r V_{\rm eff} = 0$, and $\ddot{\theta}=0$ requires $\partial_\theta V_{\rm eff} = 0$. The orbits are stable under small perturbations along the radial direction if $\partial_r^2 V_{\rm eff} > 0$, and they are stable under small perturbations along the vertical direction if $\partial_\theta^2 V_{\rm eff} > 0$.

\begin{figure*}[t]
\begin{center}
\includegraphics[type=pdf,ext=.pdf,read=.pdf,width=7.6cm]{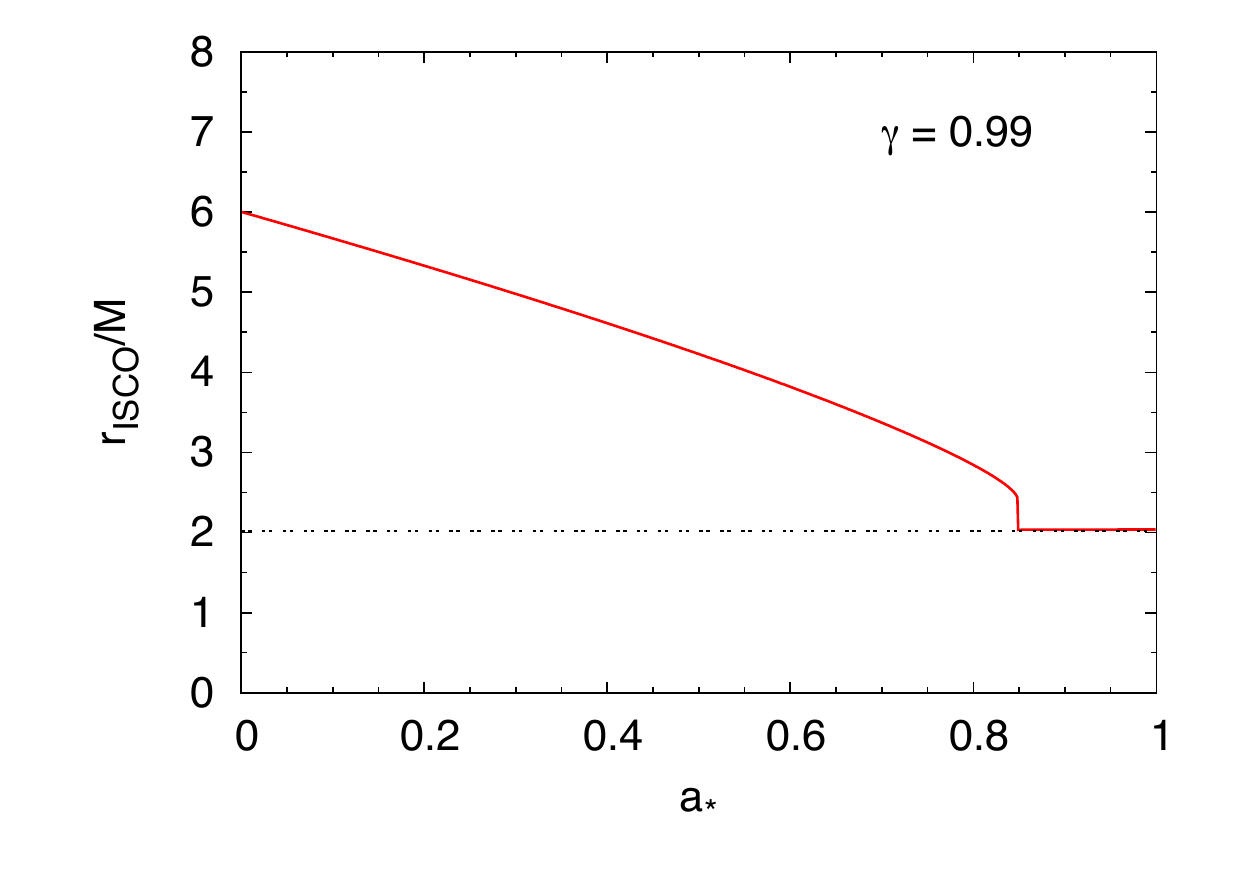}
\includegraphics[type=pdf,ext=.pdf,read=.pdf,width=7.6cm]{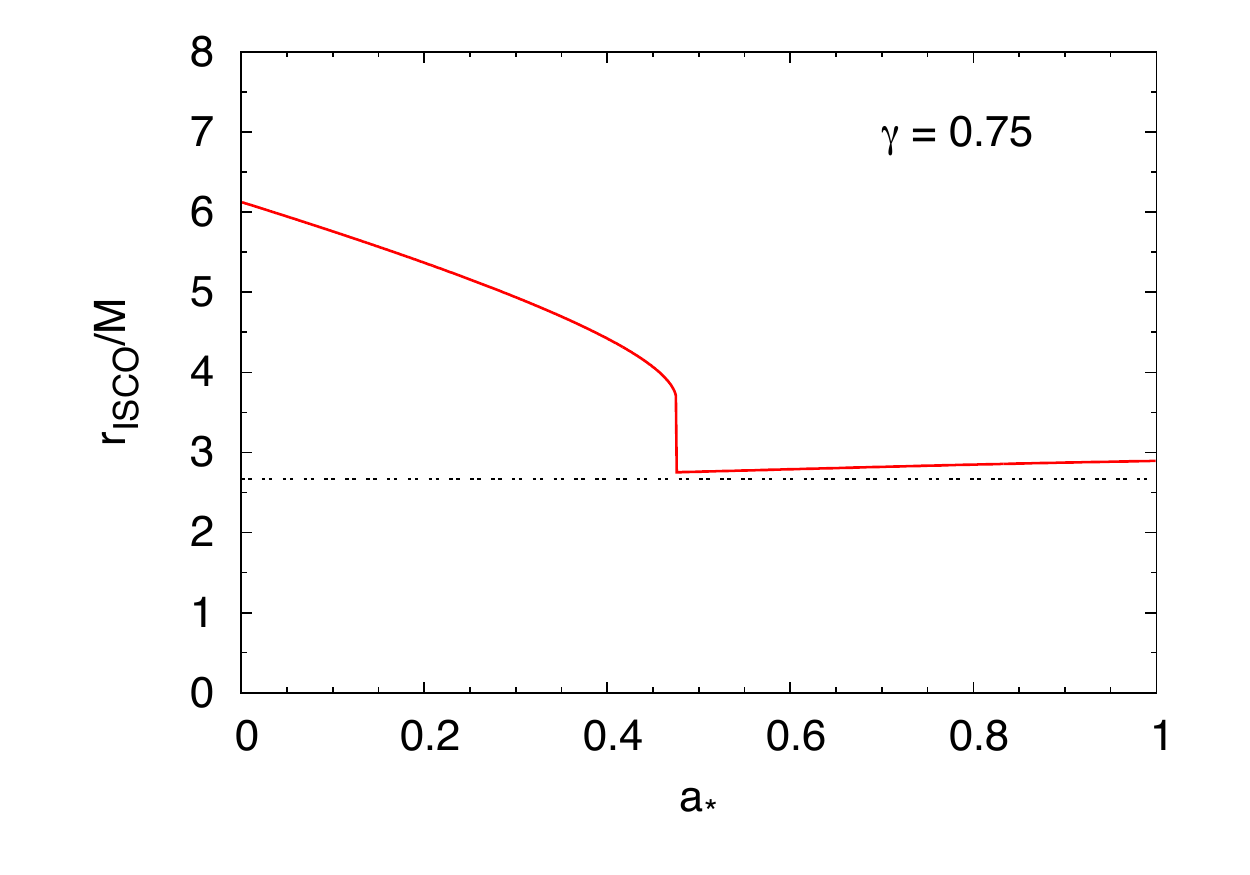} \\
\includegraphics[type=pdf,ext=.pdf,read=.pdf,width=7.6cm]{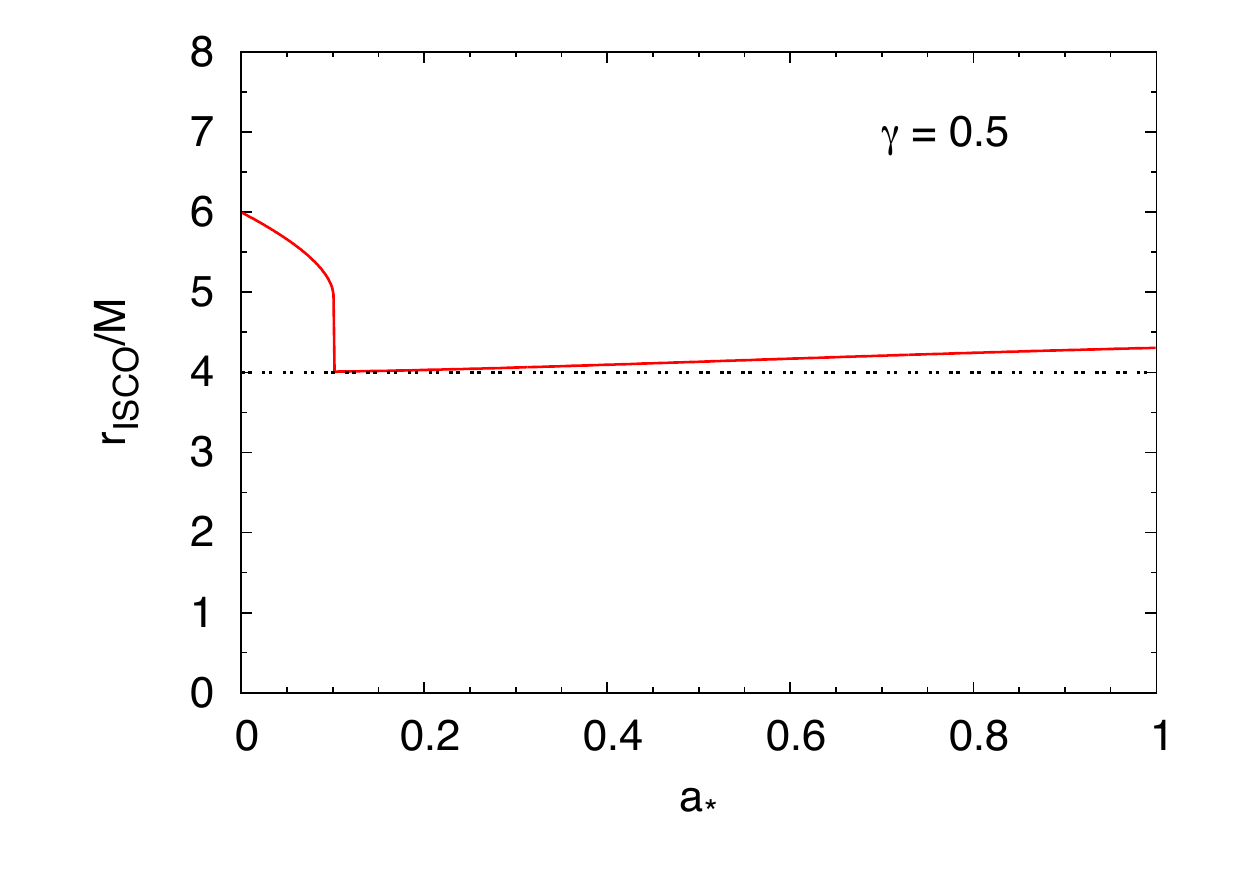}
\includegraphics[type=pdf,ext=.pdf,read=.pdf,width=7.6cm]{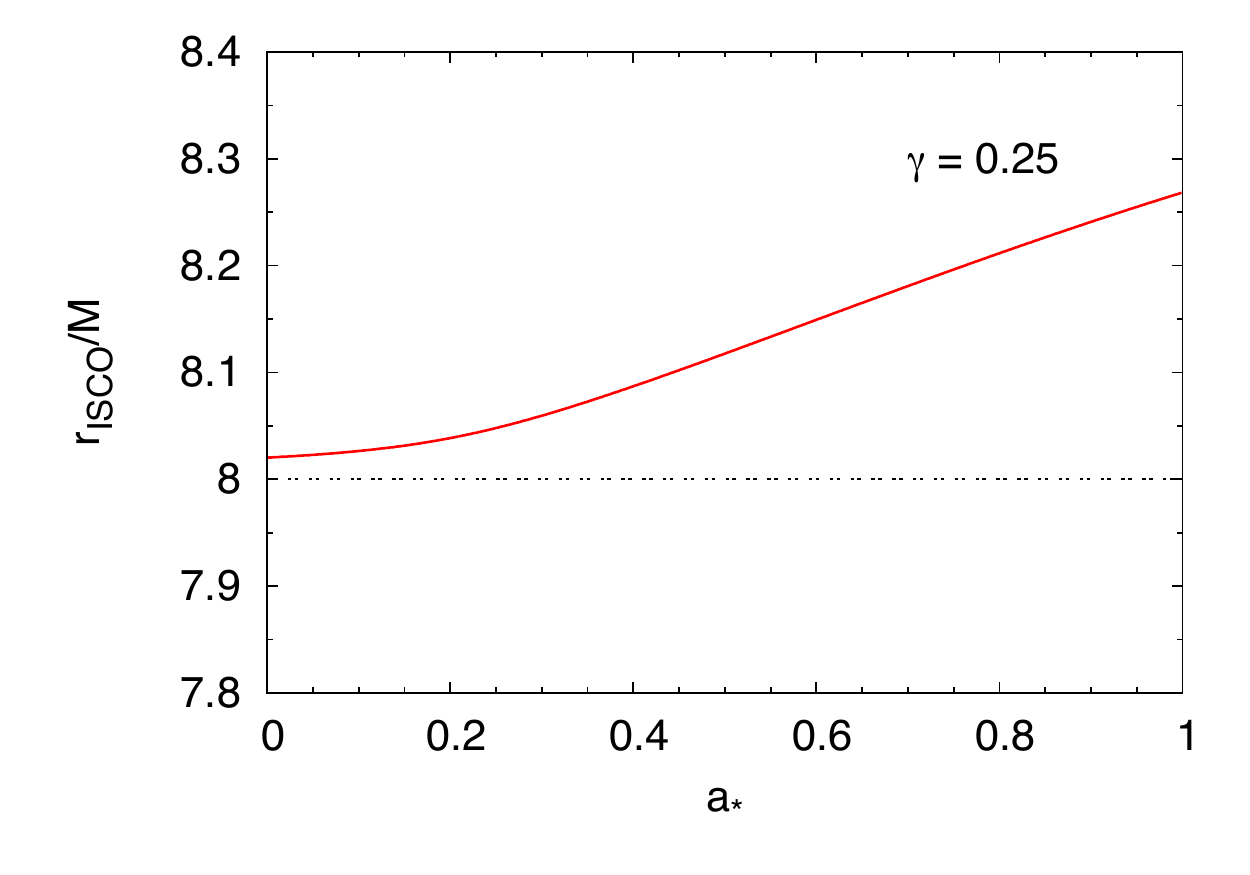}
\end{center}
\vspace{-0.5cm}
\caption{ISCO radius as a function of the dimensionless spin parameter $a_*$ for different value of the parameter $\gamma$: $\gamma = 0.99$ (top left panel), 0.75 (top right panel), 0.5 (bottom left panel), and 0.25 (bottom right panel). In every plot, the black dotted horizontal line shows the radius of the singular surface $r_*$. Note the difference in the range of $r_{\rm ISCO}$ in the bottom right panel and its large value of $r_*$. \label{f-isco}}
\end{figure*}

Fig.~\ref{f-isco} shows the ISCO radius $r_{\rm ISCO}$ as a function of the spin parameter $a_*$ for different values of the parameter $\gamma$. The black dotted horizontal lines in every plot denote the radius of the singular surface $r_*$. Since $r_*$ increases as $\gamma$ moves from 1 to 0, we can already anticipate that the gravitational redshift experienced by the photons emitted from the inner part of the accretion disk gets milder and milder. Note that the ISCO radius is determined by the radial stability of the orbit (as it is always the case in Kerr spacetimes) for ``small'' values of $a_*$, then the ISCO radius cannot be defined or drops to some radial coordinate close to $r_*$, and for larger $a_*$ the ISCO is determined by the stability of the orbit along the vertical direction.

\section{Iron line shapes \label{s-iron}}

Broad and skewed iron lines are a common feature in the X-ray spectrum of astrophysical black hole candidates. They are thought to be generated in the inner part of the accretion disk. The iron K$\alpha$ line is a very narrow feature at 6.4~keV in the case of neutral or weakly ionized iron and shifts up to 6.97~keV in the case of H-like iron ions. However, the iron lines observed in the X-ray spectra of black hole candidates are broad and skewed as the result of relativistic effects occurring in the strong gravity region (gravitational redshift, Doppler boosting, light bending). In the presence of high quality data and with the correct astrophysical model, iron line spectroscopy can be a powerful tool to test the nature of astrophysical black holes~\cite{jjc}. Reliable tests with real data require to fit the full reflection spectrum of the disk~\cite{book}, not only the iron line, but here, as a preliminary and explorative study to distinguish black holes from naked singularities, we simplify our analysis and we restrict the attention to the iron line only. This is indeed the strongest feature in the reflection spectrum and the main source of information about the spacetime geometry.

The iron line shape observed far from the source is determined by the background metric, the inclination angle of the disk, the emissivity region, and the intensity profile of the accretion disk. In our study of the JNW metric, the spacetime metric is characterized by two parameters, namely the spin parameter $a_*$ and the deformation parameter $\gamma$ related to the scalar charge of the source. The mass $M$ does not ``directly''\footnote{For simplicity, here we assume a neutral iron line at 6.4~keV. In the reality, the gas of the disk is ionized. The mass $M$ determines the effective temperature of the disk, which affects the ionization of the gas, and, consequently, the exact iron line energy.} affect the shape of the iron line because it only sets the scale of the system, while the iron line shape is determined by the redshift, namely the ratio between the photon energy at the detection and the emission points. The inclination angle of the disk is described by the angle $i$ between the spin axis and the line of sight of the observer. The emission region is the disk, with the inner edge at the ISCO radius and the outer edge at $r_{\rm ISCO} + 100 \; M$ in our calculations; the latter is large enough that its impact is small, because the emissivity at large radii is strongly suppressed. The emissivity profile is modeled with a simple power-law $1/r^q$, where $q$ is the emissivity index.

\begin{figure*}[t]
\begin{center}
\includegraphics[type=pdf,ext=.pdf,read=.pdf,width=5.4cm]{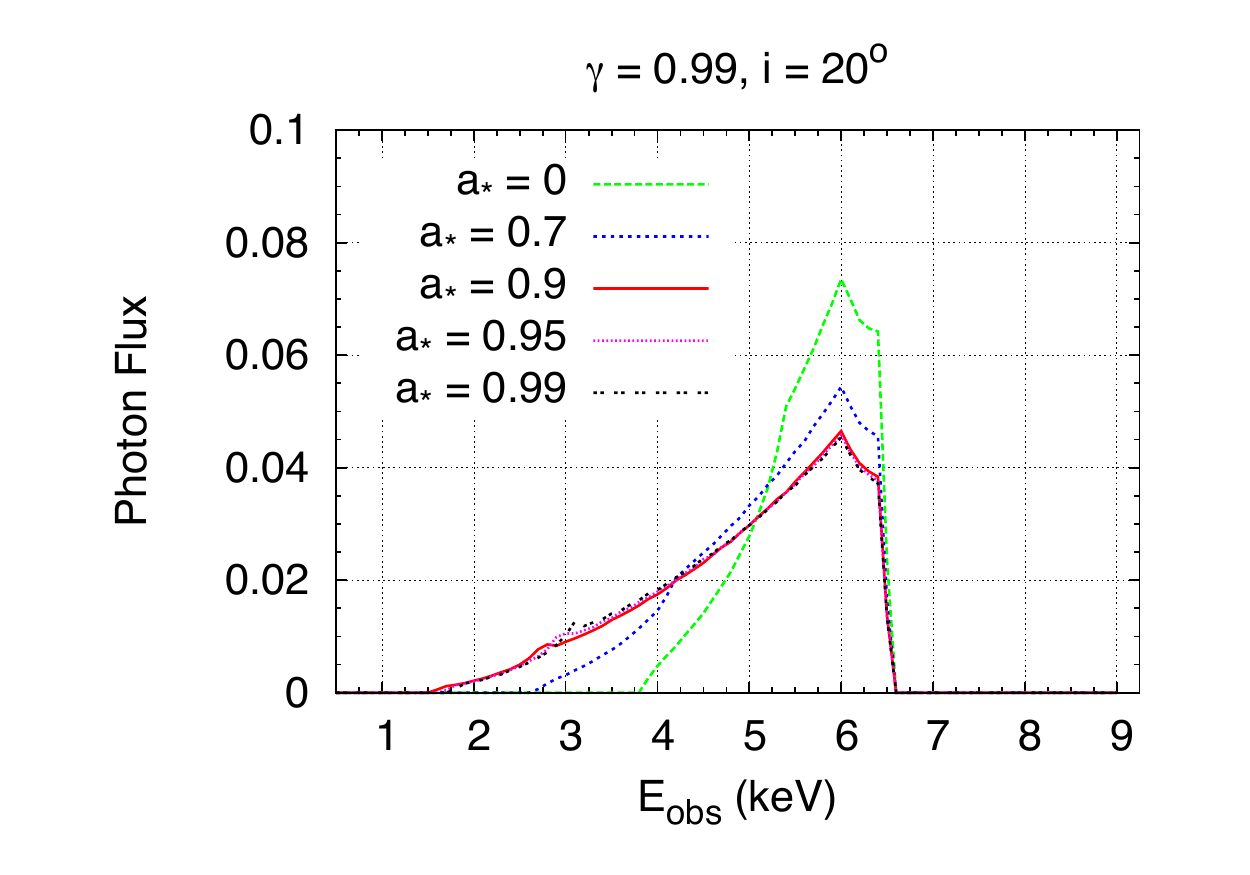}
\hspace{-0.8cm}
\includegraphics[type=pdf,ext=.pdf,read=.pdf,width=5.4cm]{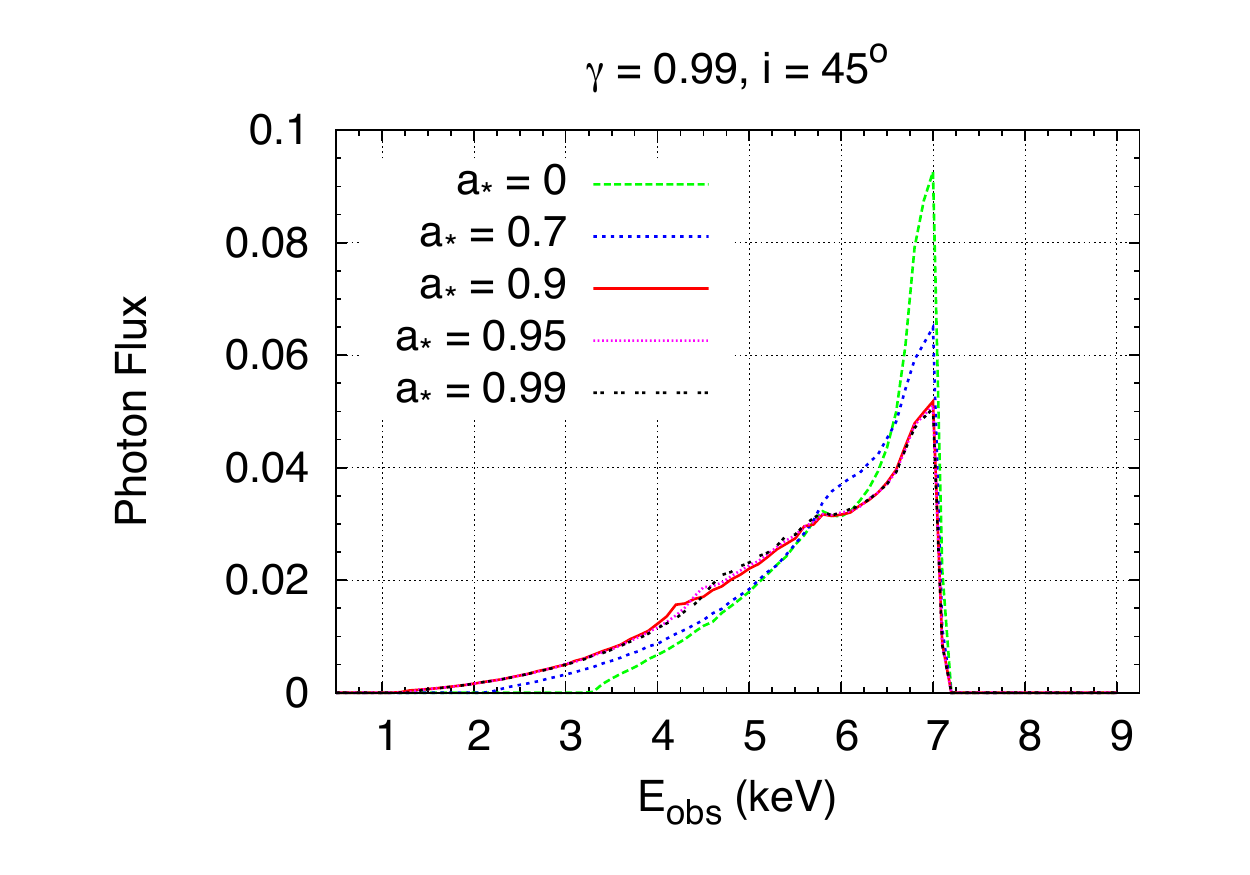}
\hspace{-0.8cm}
\includegraphics[type=pdf,ext=.pdf,read=.pdf,width=5.4cm]{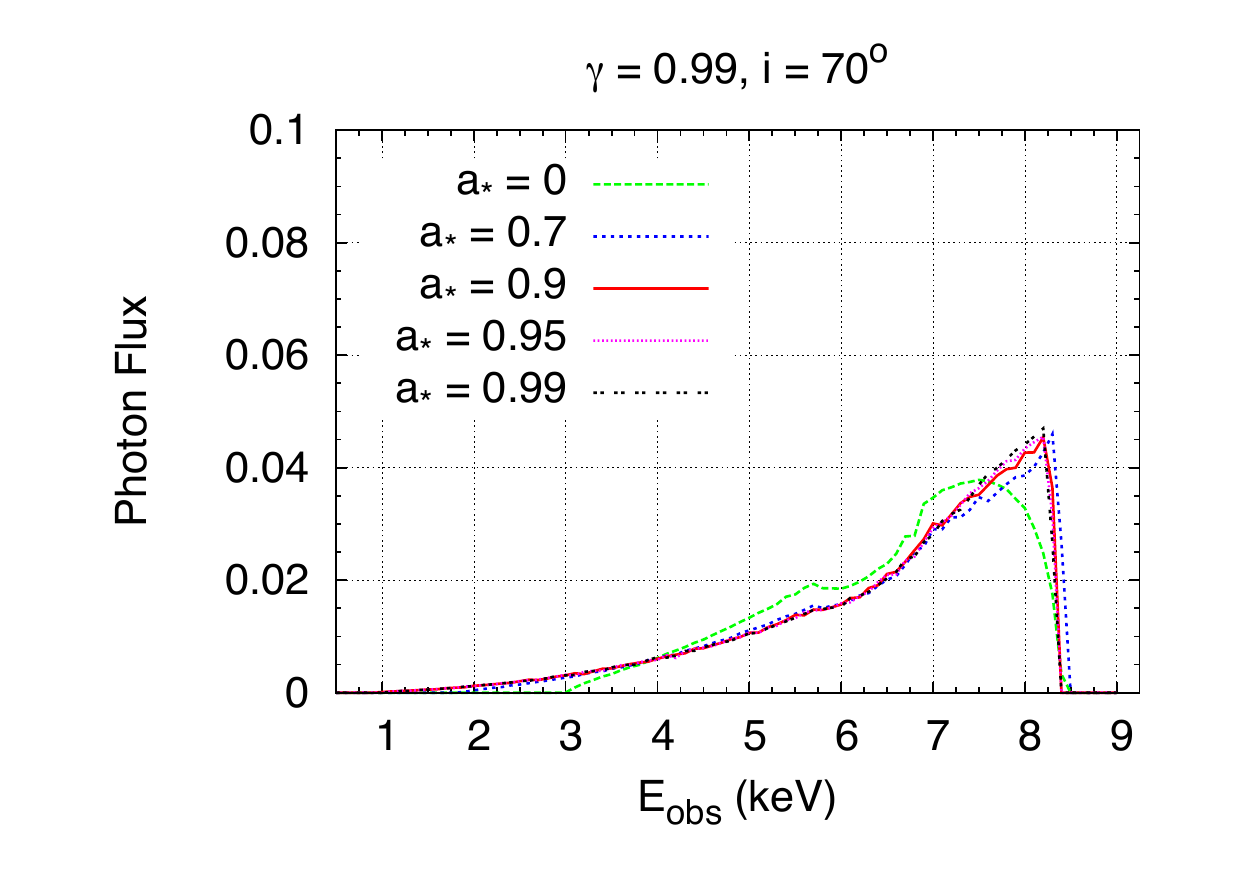} \\
\includegraphics[type=pdf,ext=.pdf,read=.pdf,width=5.4cm]{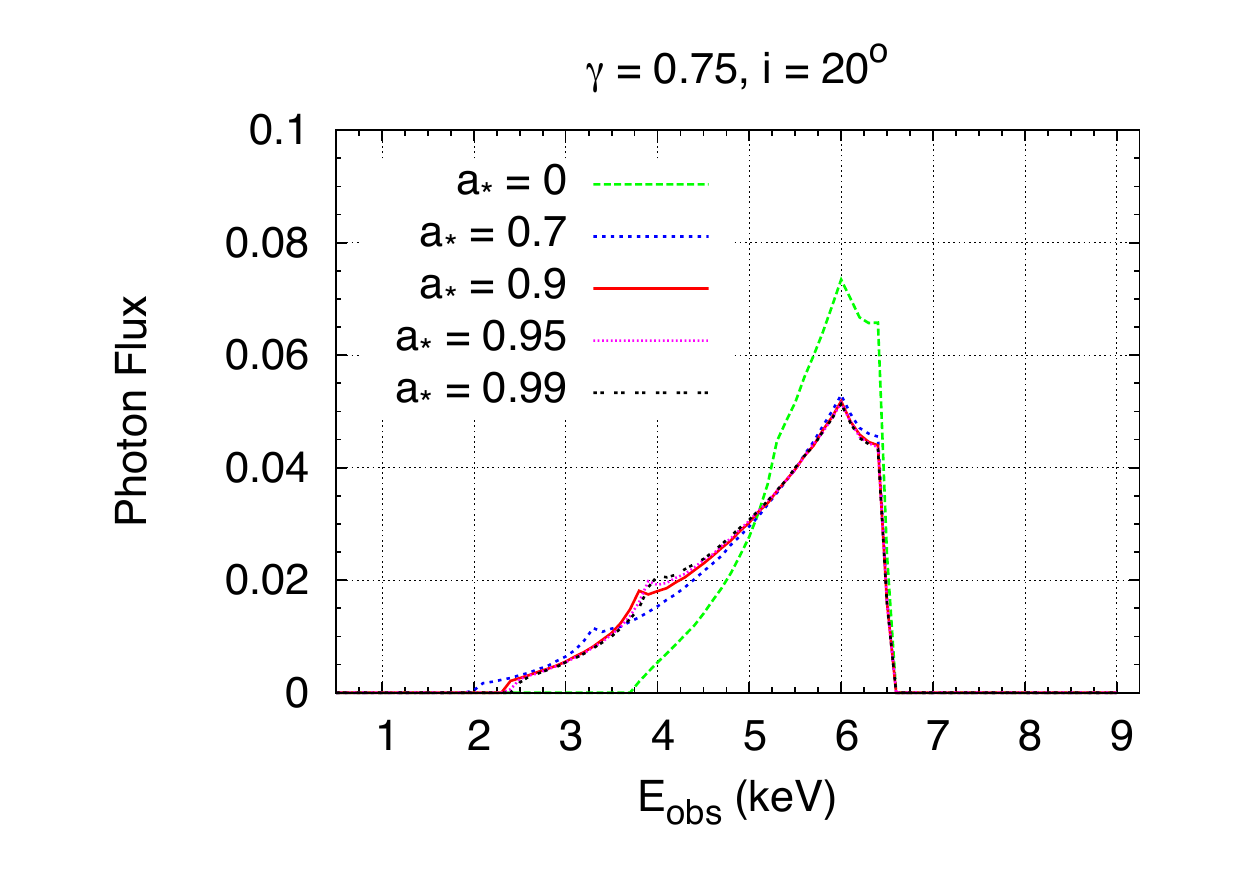}
\hspace{-0.8cm}
\includegraphics[type=pdf,ext=.pdf,read=.pdf,width=5.4cm]{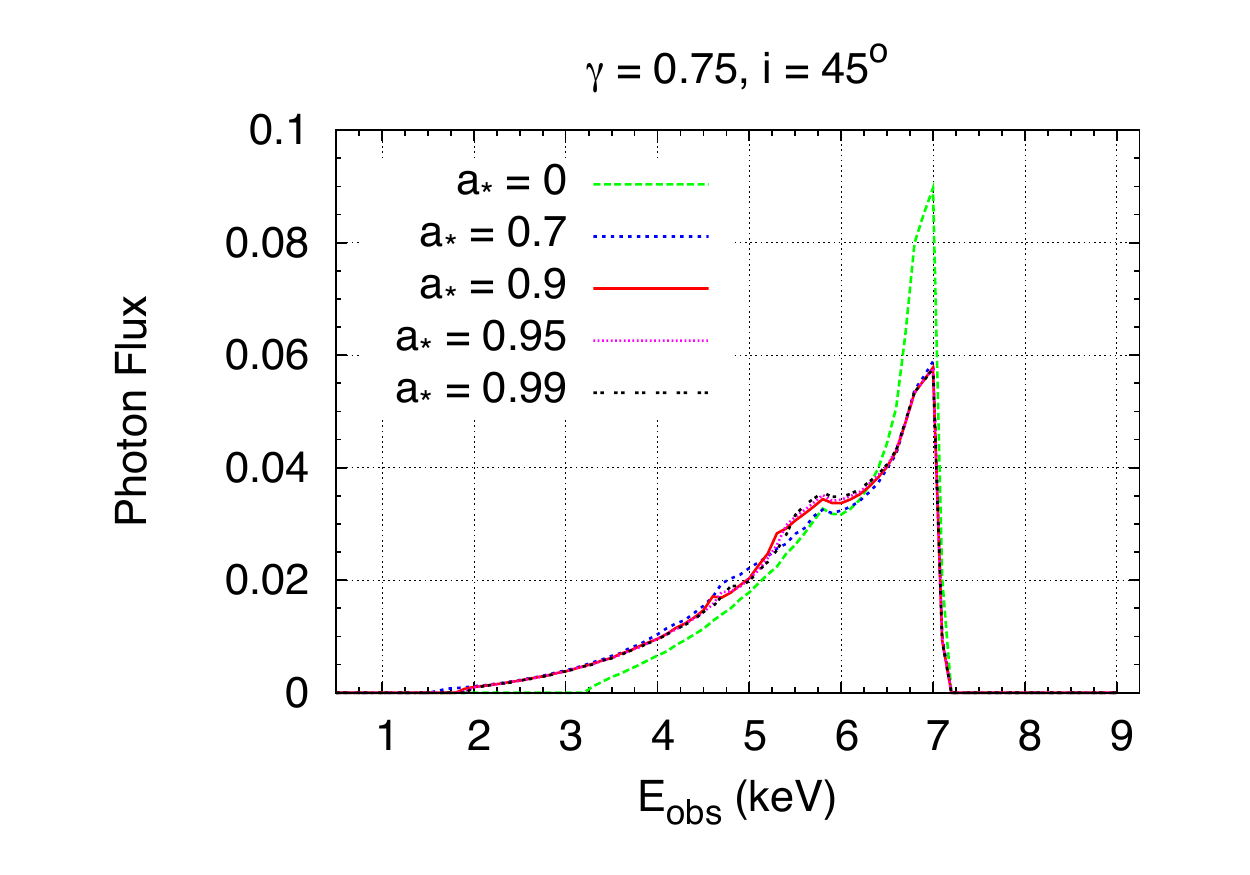}
\hspace{-0.8cm}
\includegraphics[type=pdf,ext=.pdf,read=.pdf,width=5.4cm]{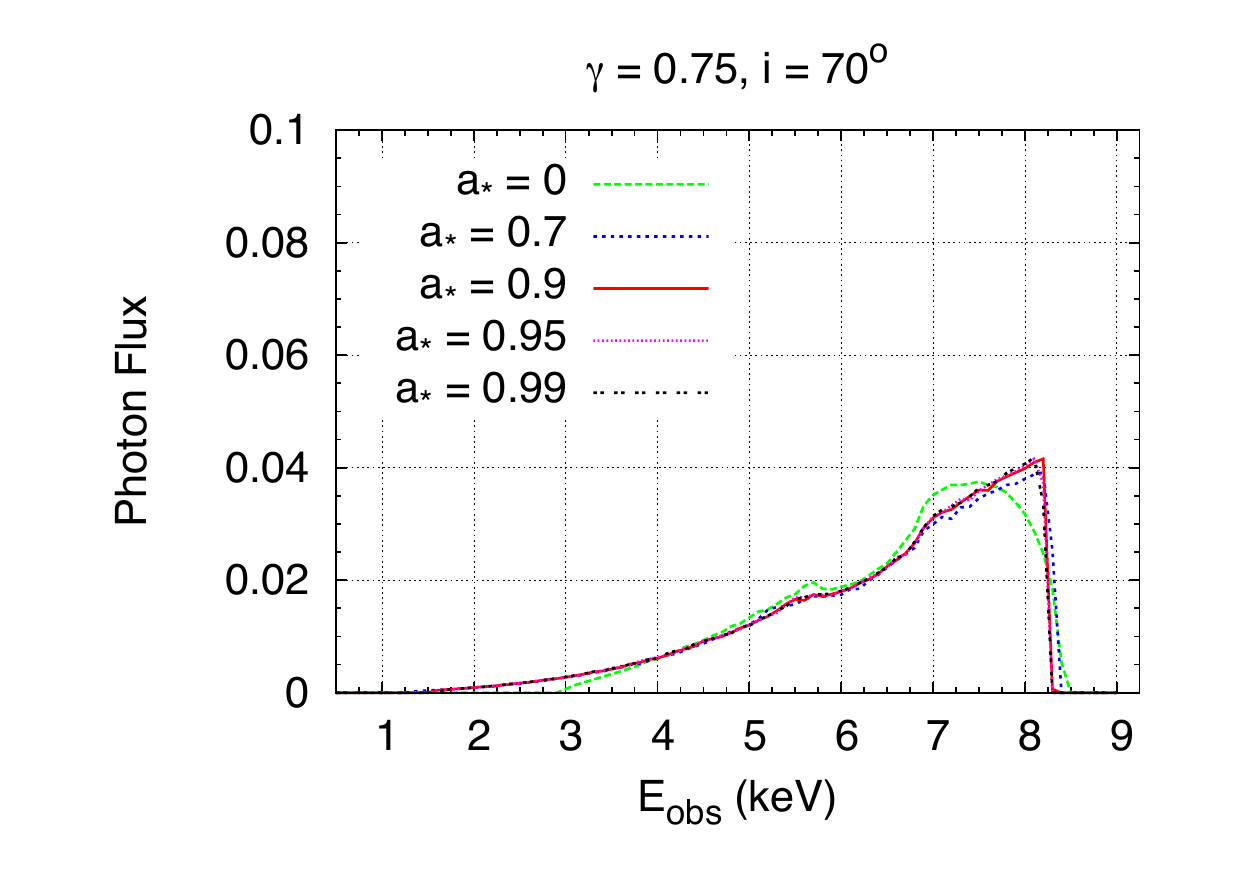} \\
\includegraphics[type=pdf,ext=.pdf,read=.pdf,width=5.4cm]{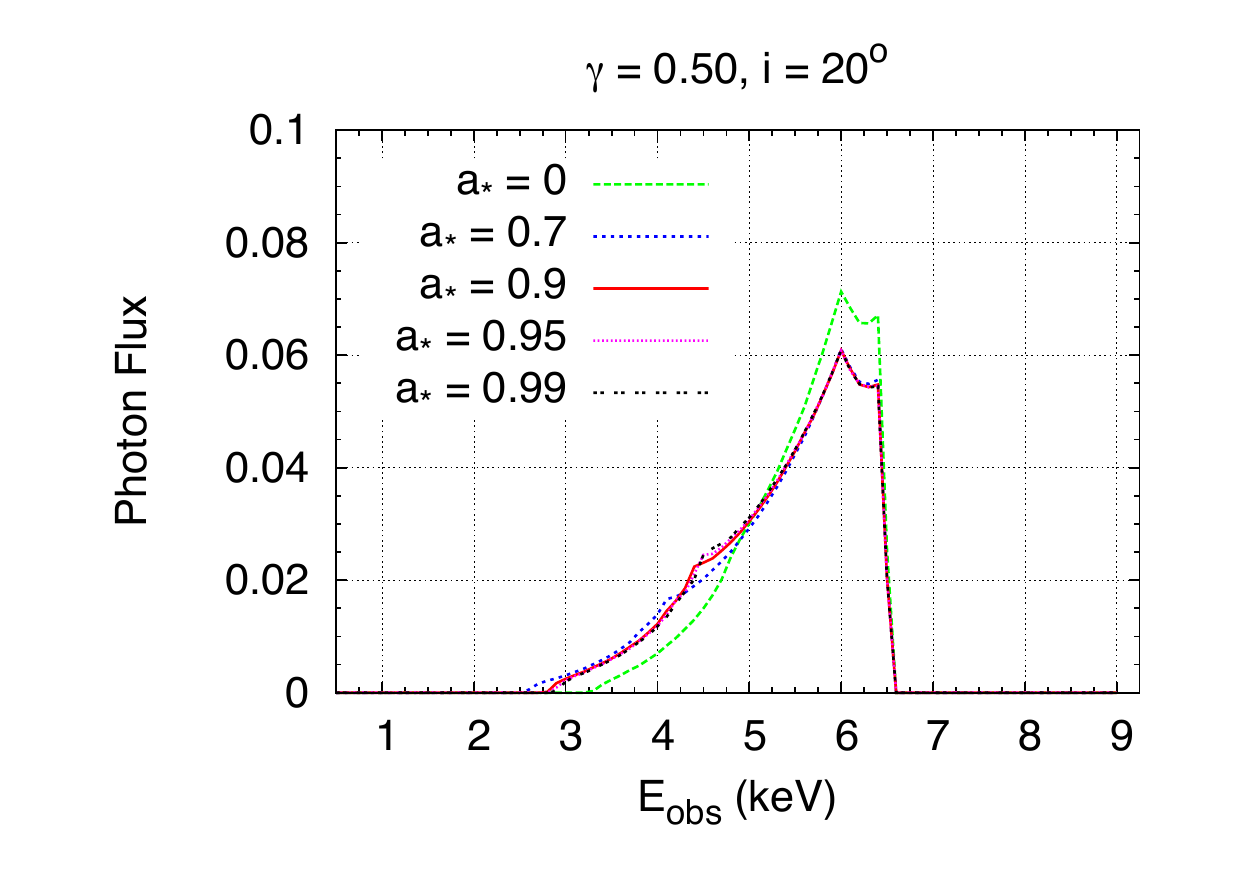}
\hspace{-0.8cm}
\includegraphics[type=pdf,ext=.pdf,read=.pdf,width=5.4cm]{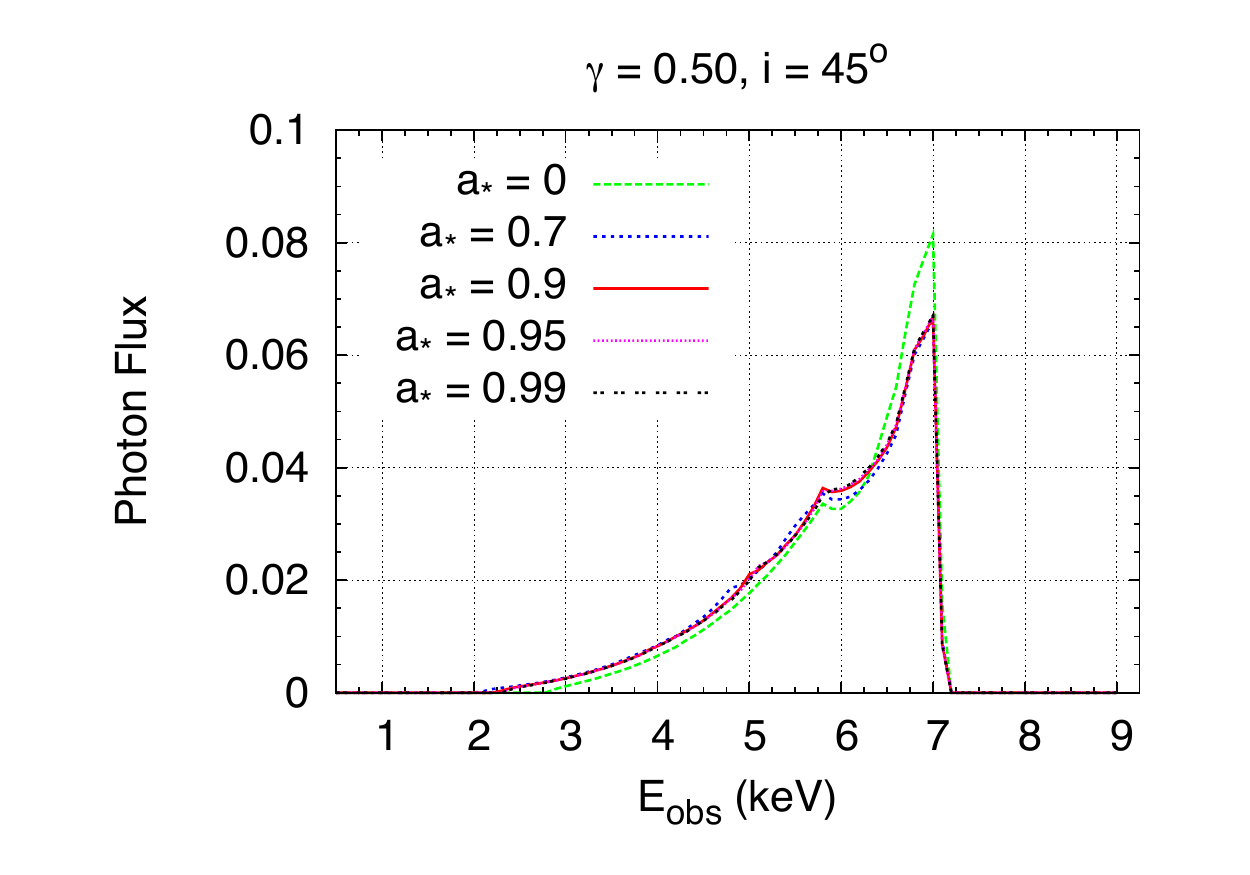}
\hspace{-0.8cm}
\includegraphics[type=pdf,ext=.pdf,read=.pdf,width=5.4cm]{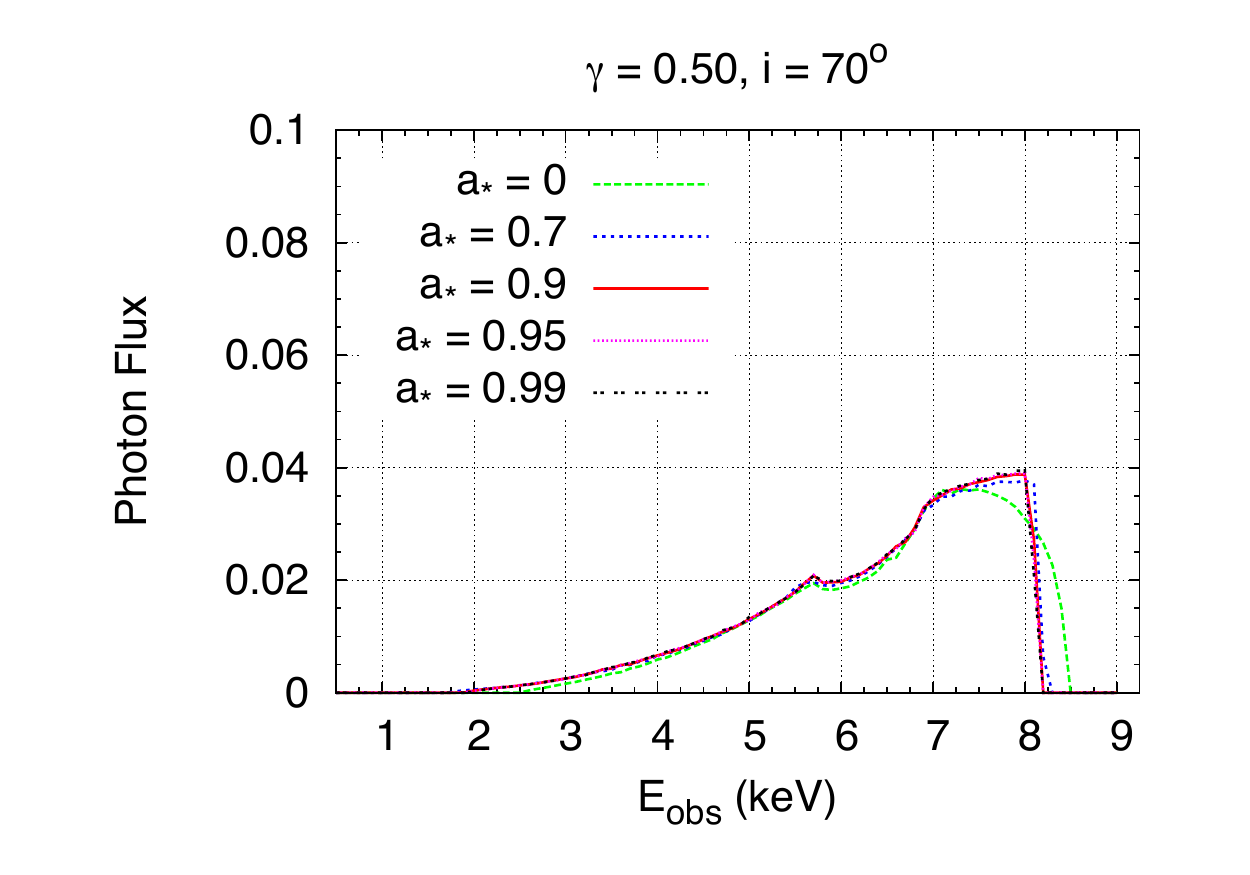} \\
\includegraphics[type=pdf,ext=.pdf,read=.pdf,width=5.4cm]{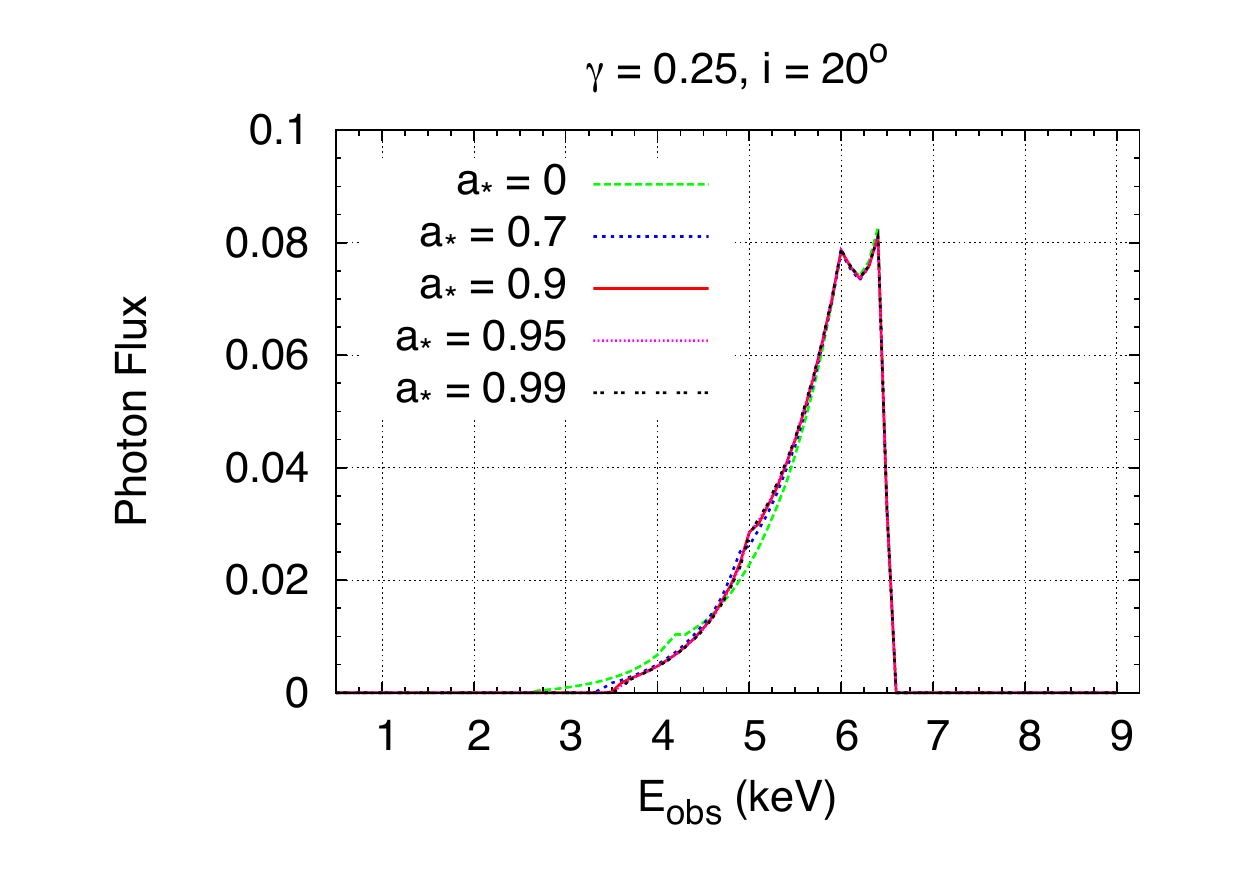}
\hspace{-0.8cm}
\includegraphics[type=pdf,ext=.pdf,read=.pdf,width=5.4cm]{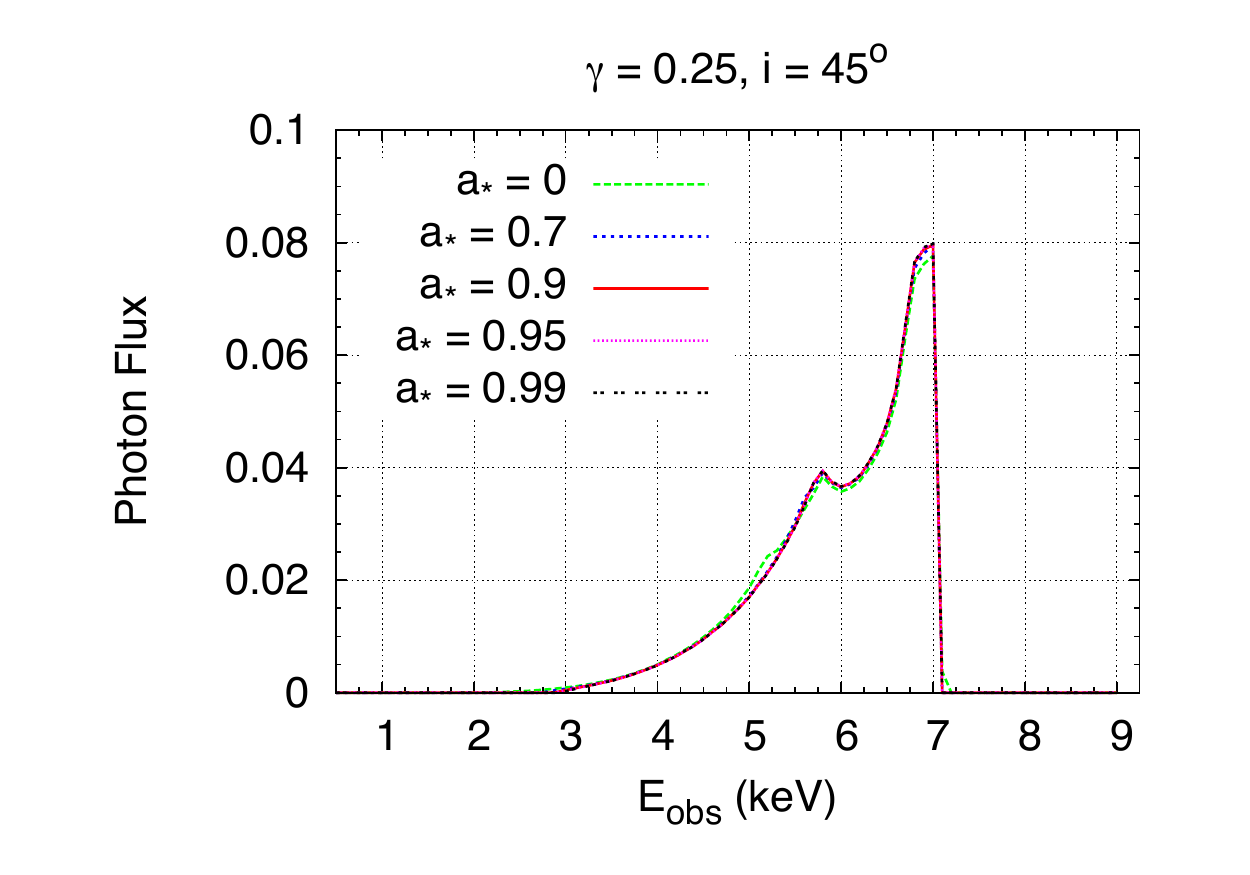}
\hspace{-0.8cm}
\includegraphics[type=pdf,ext=.pdf,read=.pdf,width=5.4cm]{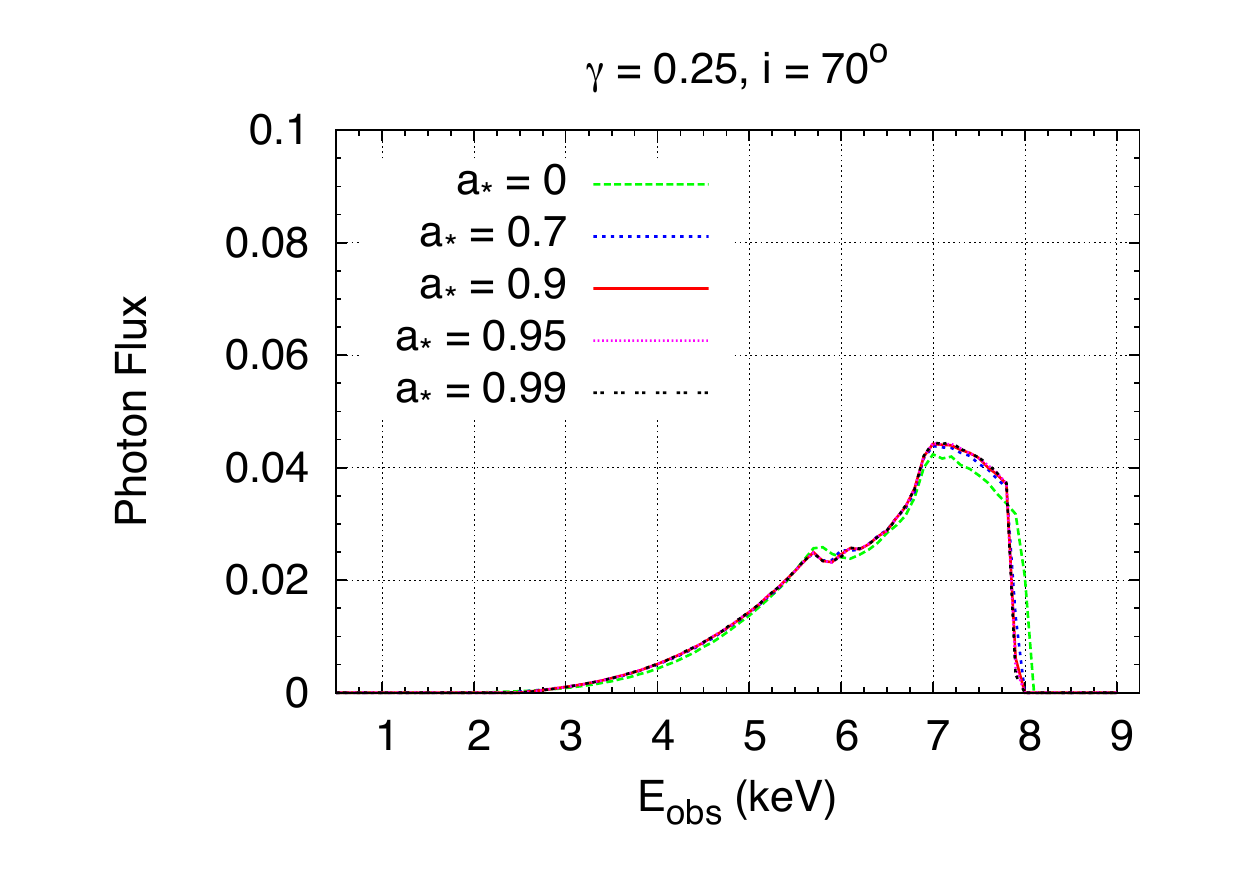}
\end{center}
\vspace{-0.5cm}
\caption{Simulations of iron line shapes in the reflection spectrum of an accretion disk around JNW naked singularities. The parameter $\gamma$ is set at 0.99, 0.75, 0.50, and 0.25 (from top to bottom). The inclination angle of the disk with respect to the line of sight of the distant observer is $i = 20^\circ$ (left panels), $45^\circ$ (central panels), and $70^\circ$ (right panels). The rest-frame energy of the iron line is 6.4~keV and the emissivity profile is a power law with emissivity index 3. \label{f-iron}}
\end{figure*}

Iron line shapes in the JNW spacetime have been calculated and reported in Fig.~\ref{f-iron} for different values of $a_*$, $\gamma$, and $i$. The calculations have been done with the code described in Refs.~\cite{code1,code2}, and the method is briefly reviewed in Appendix~\ref{s-app}. In Fig.~\ref{f-iron}, we show the iron line for $\gamma = 0.99$, 0.75, 0.50, and 0.25 (from top to bottom). The inclination angle of the disk is $i = 20^\circ$ (left panels), $45^\circ$ (central panels), and $70^\circ$ (right panels). Every panel shows the iron line shapes for the spin parameters $a_* = 0$, 0.7, 0.9, 0.95, and 0.99. In all our simulations we have assumed the emissivity index $q=3$, which corresponds to the Newtonian case for a corona with lamppost geometry at large radii; see, e.g., Ref.~\cite{book} for more details.


\section{Simulations \label{s-sim}}

The iron line shapes calculated in the previous section do not show any very peculiar feature with respect to those in a Kerr spacetime. However, we can note that for $\gamma \rar 0$ the radius of the source $r_*$ increases, so the body becomes less and less compact, with the obvious result that the gravitational field near its surface gets weaker and the implication that the gravitational redshift experienced by photons gets milder and milder. The final result is that board iron lines are impossible for low values of $\gamma$. On the contrary, broad iron lines are observed in the X-ray spectra of many black hole candidates and are possible in the spacetime around fast-rotating Kerr black holes. Such an observation can already {\it suggest} that the iron line shape can potentially constrain the parameter $\gamma$.

To take a step forward, we follow the approach of Refs.~\cite{sim1,sim2,sim3,sim4} and we simulate some observations with a current X-ray mission. The simulated data are then analyzed with a Kerr model for the iron line to see whether the Kerr model can well fit the data or not. Note that current iron lines in the spectra of astrophysical black hole candidates are commonly fitted with iron lines calculated in the Kerr spacetime and there is no tension between data and theoretical models. Simulations and fits are done with the publicly available X-ray package XSPEC~\cite{xspecc}. As a current X-ray mission, we consider NuSTAR. The response files of its instruments can be downloaded from the NuSTAR website~\cite{nustar}. As a source, we do not consider any specific object, but we employ the typical parameters for a bright stellar-mass black hole candidate. The spectrum of the source is modeled with a power-law component ($E^{-\Gamma}$ representing the spectrum of the corona) and a single iron line (representing the reflection spectrum of the disk); see e.g. Ref.~\cite{book} for more details. The luminosity of the source is assumed to be $10^{-9}$~erg/s/cm$^2$ in the energy band 2-10~keV. The equivalent width of the iron line is set at $\sim 200$~eV. For the photon index of the power-law component, we choose $\Gamma = 1.6$. The exposure time of the observation is 200~ks and we combine the data of the two instruments on board of NuSTAR, FPMA and FPMB.

\begin{table*}[t]
\centering
\begin{tabular}{lccc|ccc}
\hline\hline
Simulation & 1 & 2 & 3 & 4 & 5 & 6 \\
Input values&&&&&& \\
$a_*$ & 0.99 & 0.99 & 0.99 & 0.99 & 0.99 & 0.99 \\
$\gamma$ & 0.99 & 0.99 & 0.99 & 0.75 & 0.75 & 0.75 \\
$i$ [deg] & 20 & 45 & 70 & 20 & 45 & 70 \\
\hline
Best-fit &&&&&& \\
$a_*$ & $0.63 \pm 0.08$ & $0.90 \pm 0.6$ & $0.93 \pm 0.02$ & $0.69 \pm 0.07$ & $0.83 \pm 0.03$ & $> 0.94$ \\
$q$ & $3.12 \pm 0.05$ & $2.96 \pm 0.13$ & $3.1 \pm 0.2$ & $3.10 \pm 0.05$ & $3.40 \pm 0.17$ & $3.0 \pm 0.2$ \\
$i$ [deg] & $18.2 \pm 1.2$ & $45.1 \pm 0.5$ & $70.3 \pm 0.3$ & $19.1 \pm 1.2$ & $45.3 \pm 0.5$ &$69.0 \pm 0.4$ \\
$\chi^2_{\rm red, min}$ & 1.08 & 1.00 & 1.06 & 1.07 & 1.06 & 1.07 \\
\hline\hline
\end{tabular}
\caption{Summary of the best-fit values for the simulations 1-6. In the fits, we have six free parameters: $a_*$, $i$, $q$, $\Gamma$, and the normalizations of the power-law continuum and of the iron line. See the text for more details. \label{t-fit1}}
\vspace{1.0cm}
\centering
\begin{tabular}{lccc|ccc}
\hline\hline
Simulation & 7 & 8 & 9 & 10 & 11 & 12 \\
Input values&&&&&& \\
$a_*$ & 0.99 & 0.99 & 0.99 & 0.99 & 0.99 & 0.99 \\
$\gamma$ & 0.5 & 0.5 & 0.5 & 0.25 & 0.25 & 0.25 \\
$i$ [deg] & 20 & 45 & 70 & 20 & 45 & 70 \\
\hline
Best-fit &&&&&& \\
$a_*$ & $0.50 \pm 0.08$ & $0.82 \pm 0.10$ & $> 0.89$ & $- 0.18 \pm 0.13$ & $< 0.07$ & $> 0.87$ \\
$q$ & $2.98 \pm 0.04$ & $2.8 \pm 0.3$ & $2.73 \pm 0.10$ & $3.05 \pm 0.09$ & $2.92 \pm 0.14$ & $2.72 \pm 0.08$ \\
$i$ [deg] & $17.8 \pm 1.0$ & $46.7 \pm 0.5$ & $66.9 \pm 0.4$ & $21.5 \pm 1.0$ & $43.5 \pm 0.6$ &$62.2 \pm 0.4$ \\
$\chi^2_{\rm red, min}$ & 1.04 & 1.05 & 1.04 & 1.06 & 1.05 & 1.08 \\
\hline\hline
\end{tabular}
\caption{As in Tab.~\ref{t-fit1} for the simulations 7-12. See the text for more details. \label{t-fit2}}
\end{table*}

\begin{figure*}[t]
\begin{center}
\includegraphics[type=pdf,ext=.pdf,read=.pdf,width=7.9cm]{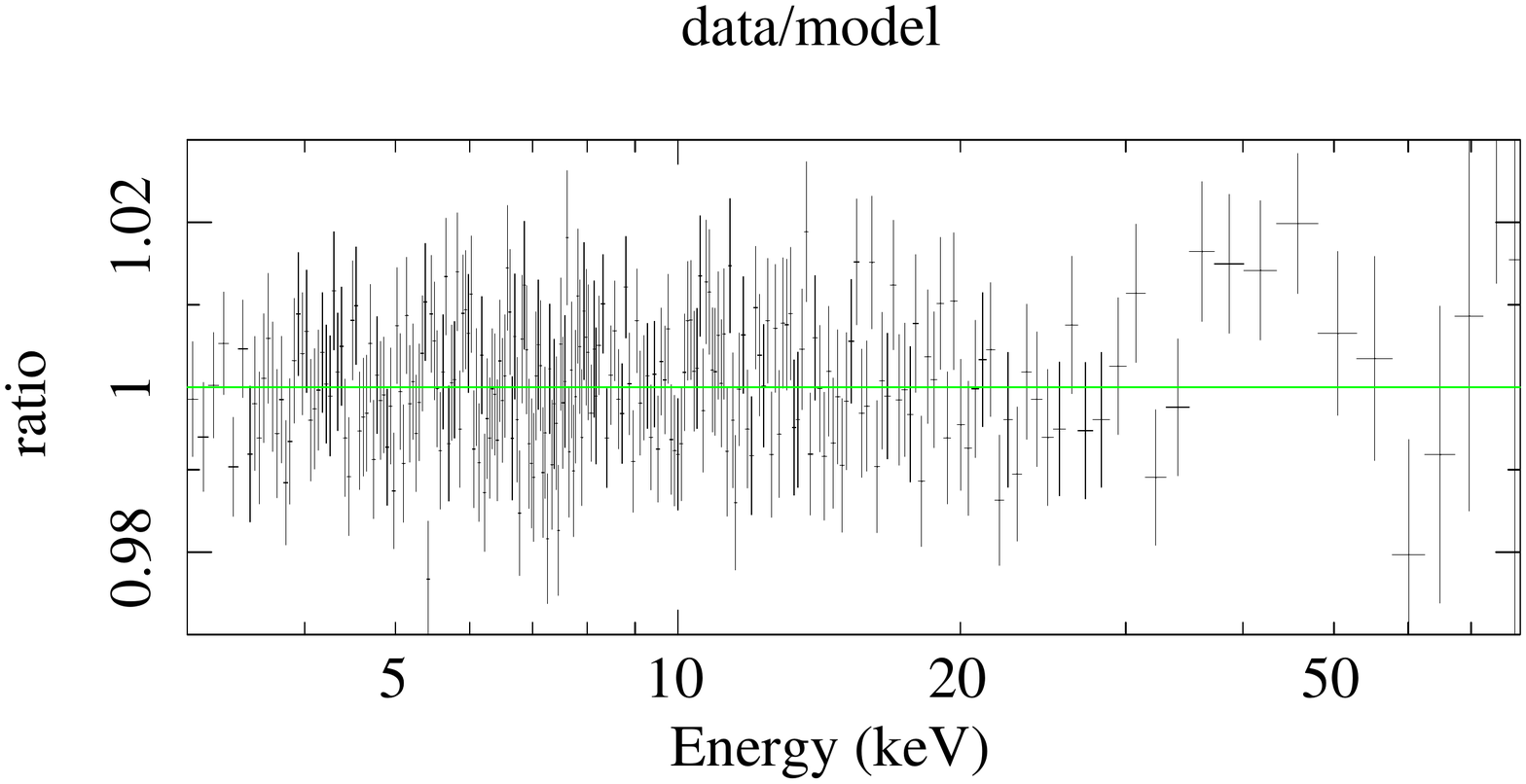} 
\hspace{-0.8cm}
\includegraphics[type=pdf,ext=.pdf,read=.pdf,width=7.9cm]{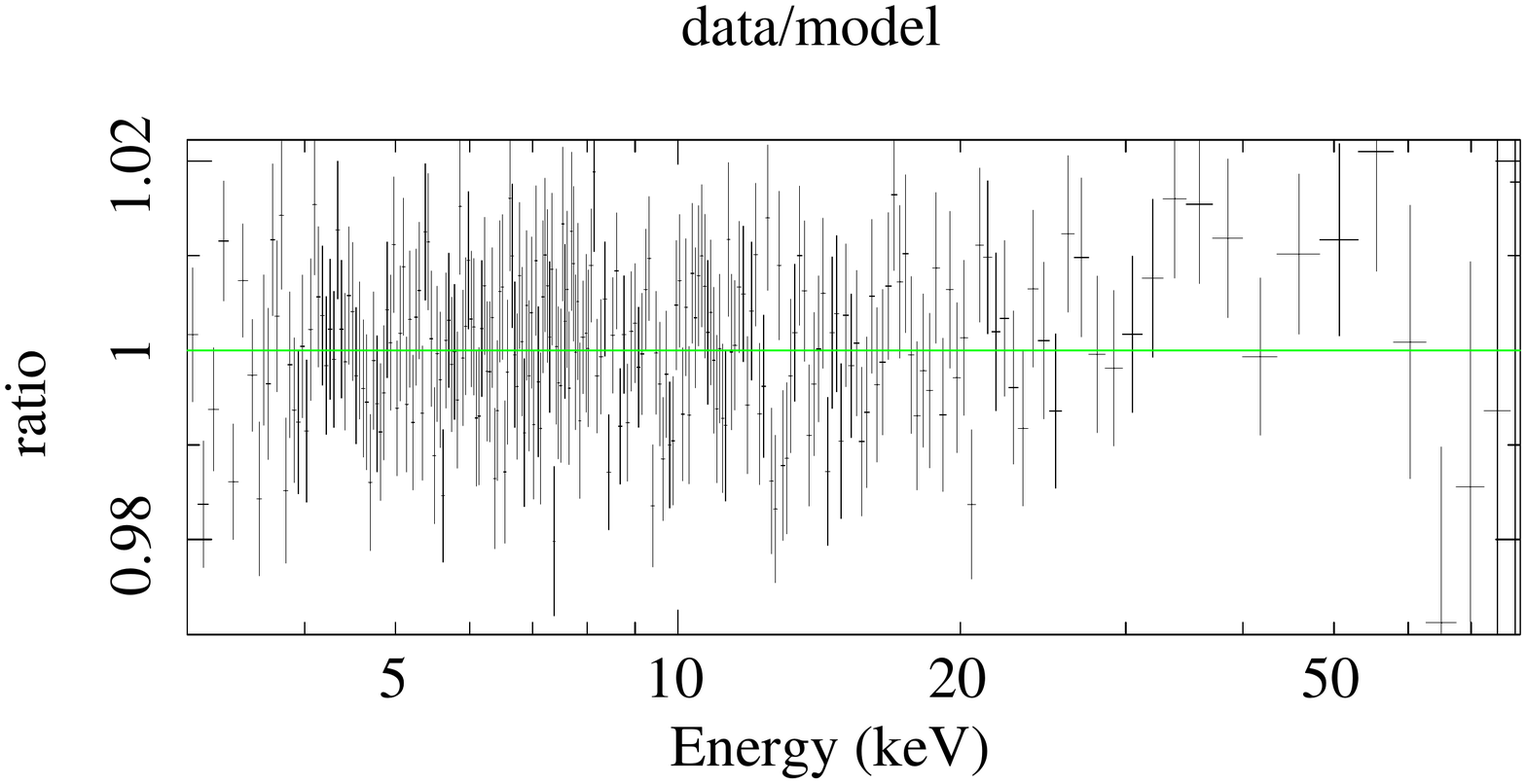}\\
\vspace{-1.9cm}
\includegraphics[type=pdf,ext=.pdf,read=.pdf,width=7.9cm]{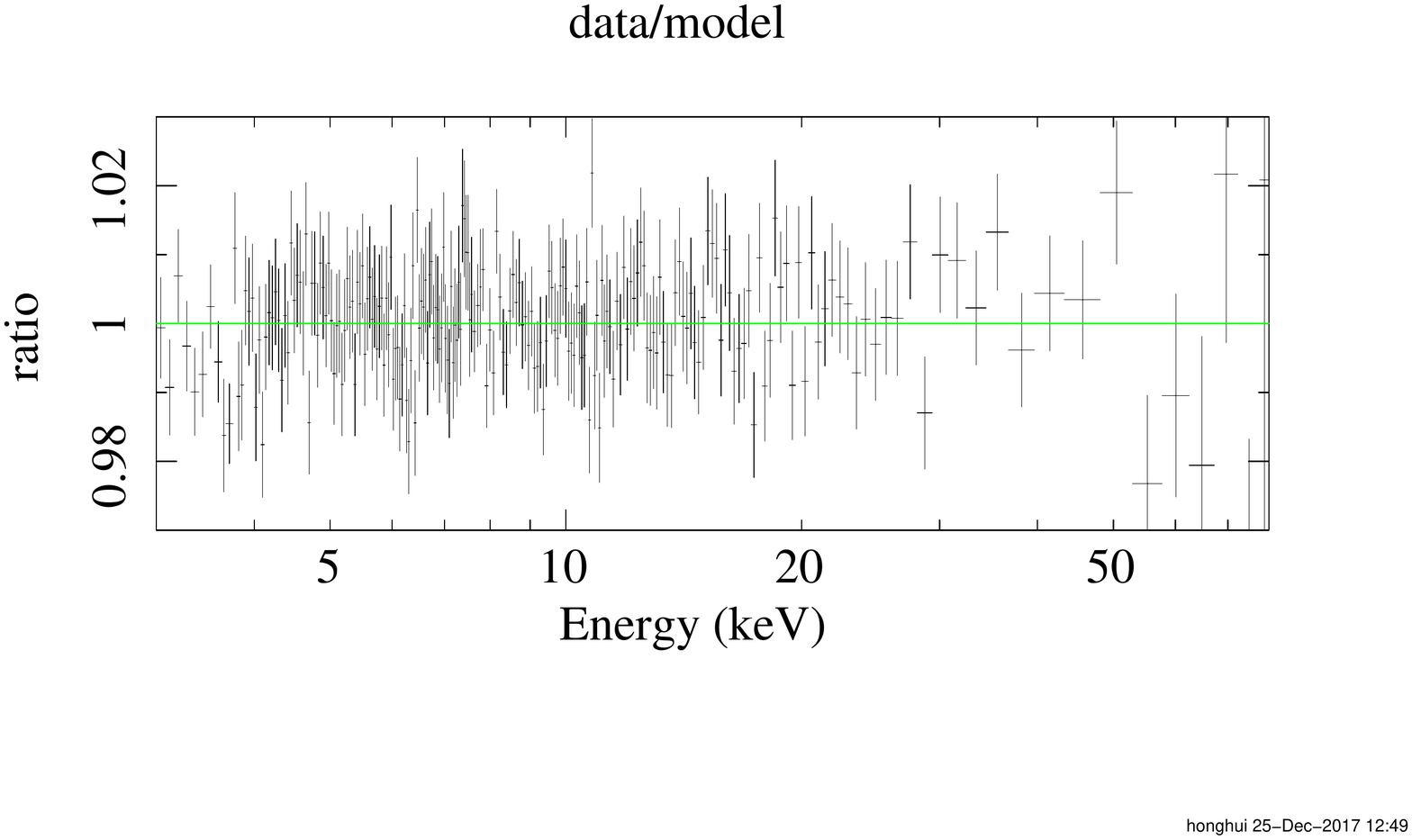}
\hspace{-0.8cm}
\includegraphics[type=pdf,ext=.pdf,read=.pdf,width=7.9cm]{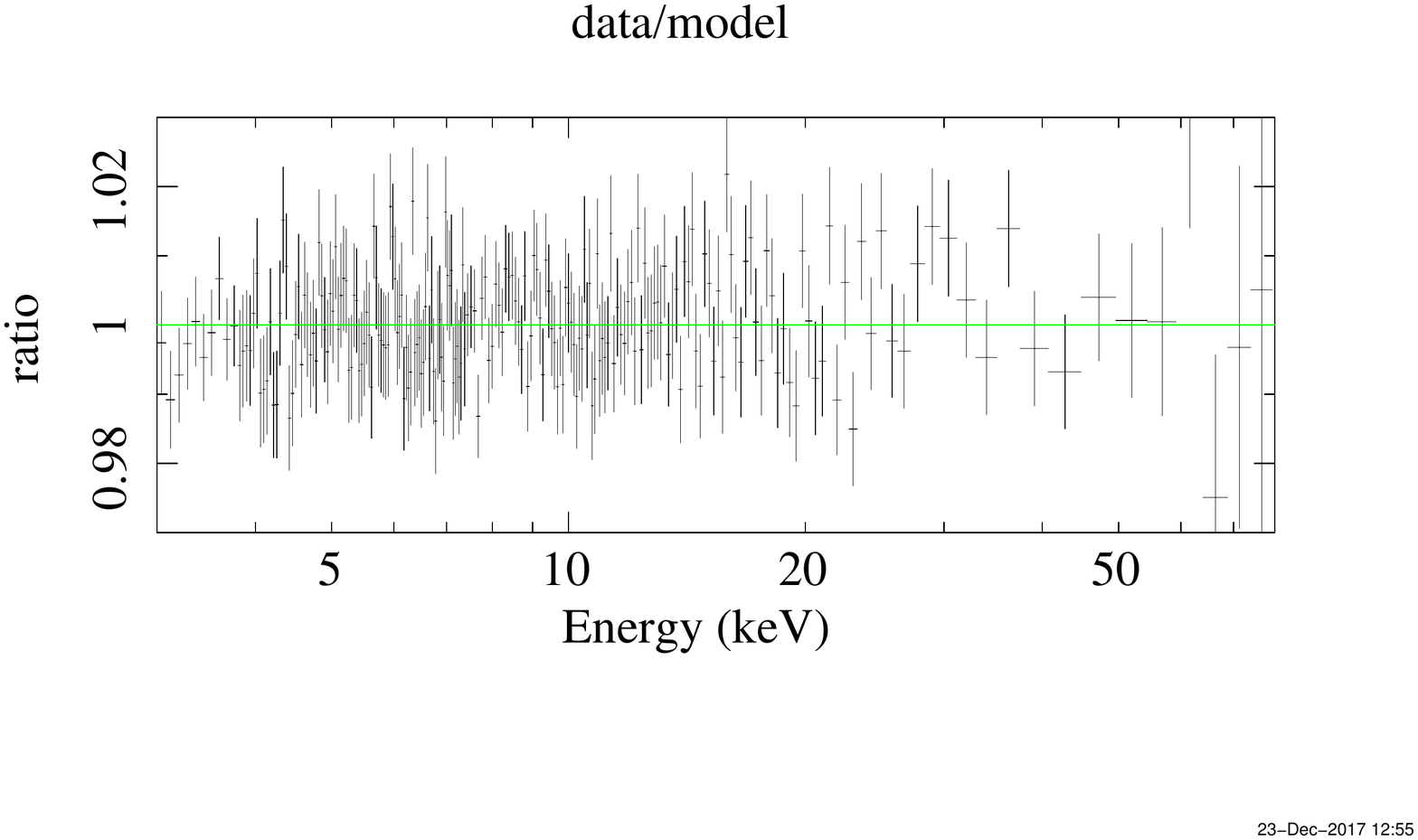}\\
\vspace{-1.9cm}
\includegraphics[type=pdf,ext=.pdf,read=.pdf,width=7.9cm]{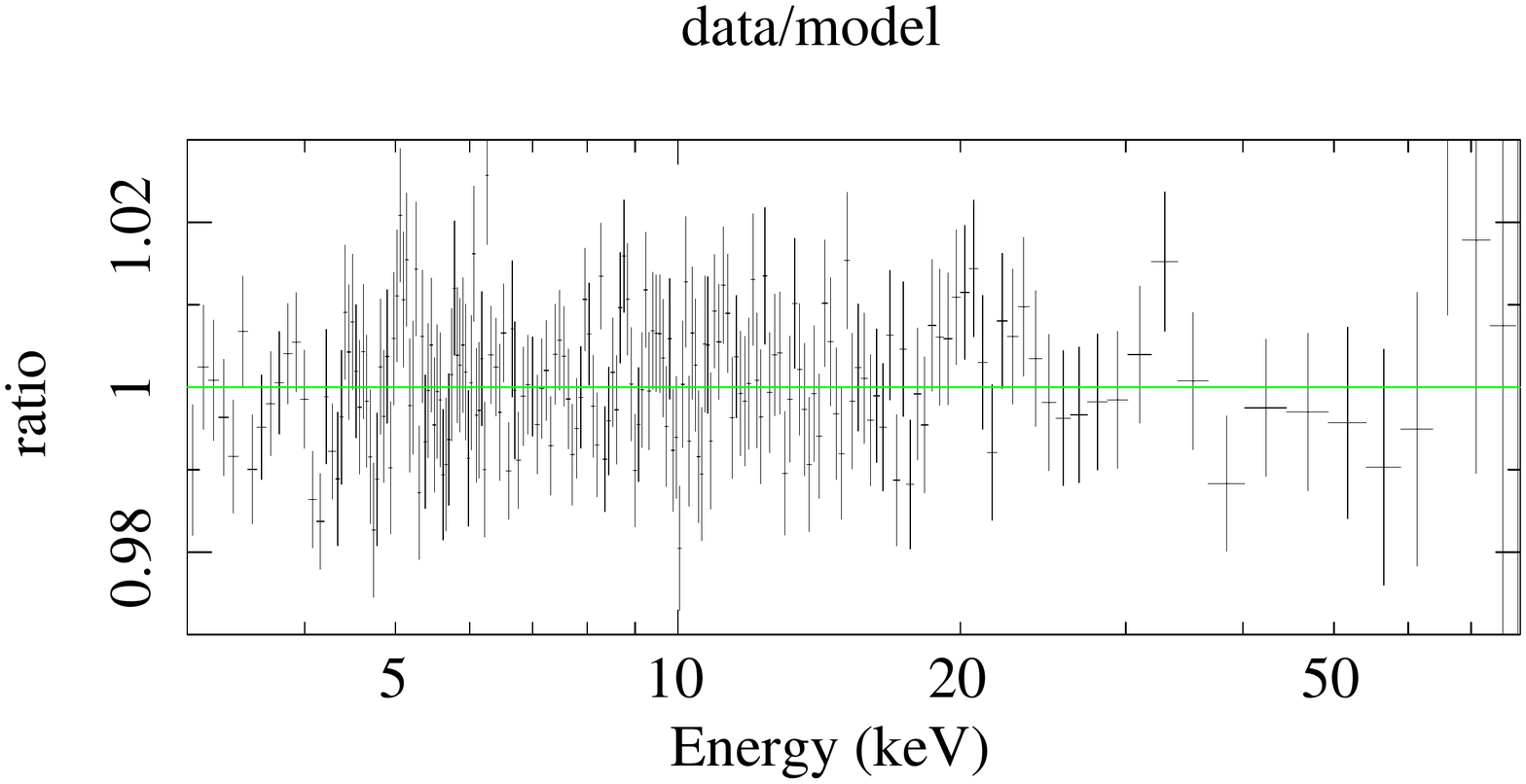}
\hspace{-0.8cm}
\includegraphics[type=pdf,ext=.pdf,read=.pdf,width=7.9cm]{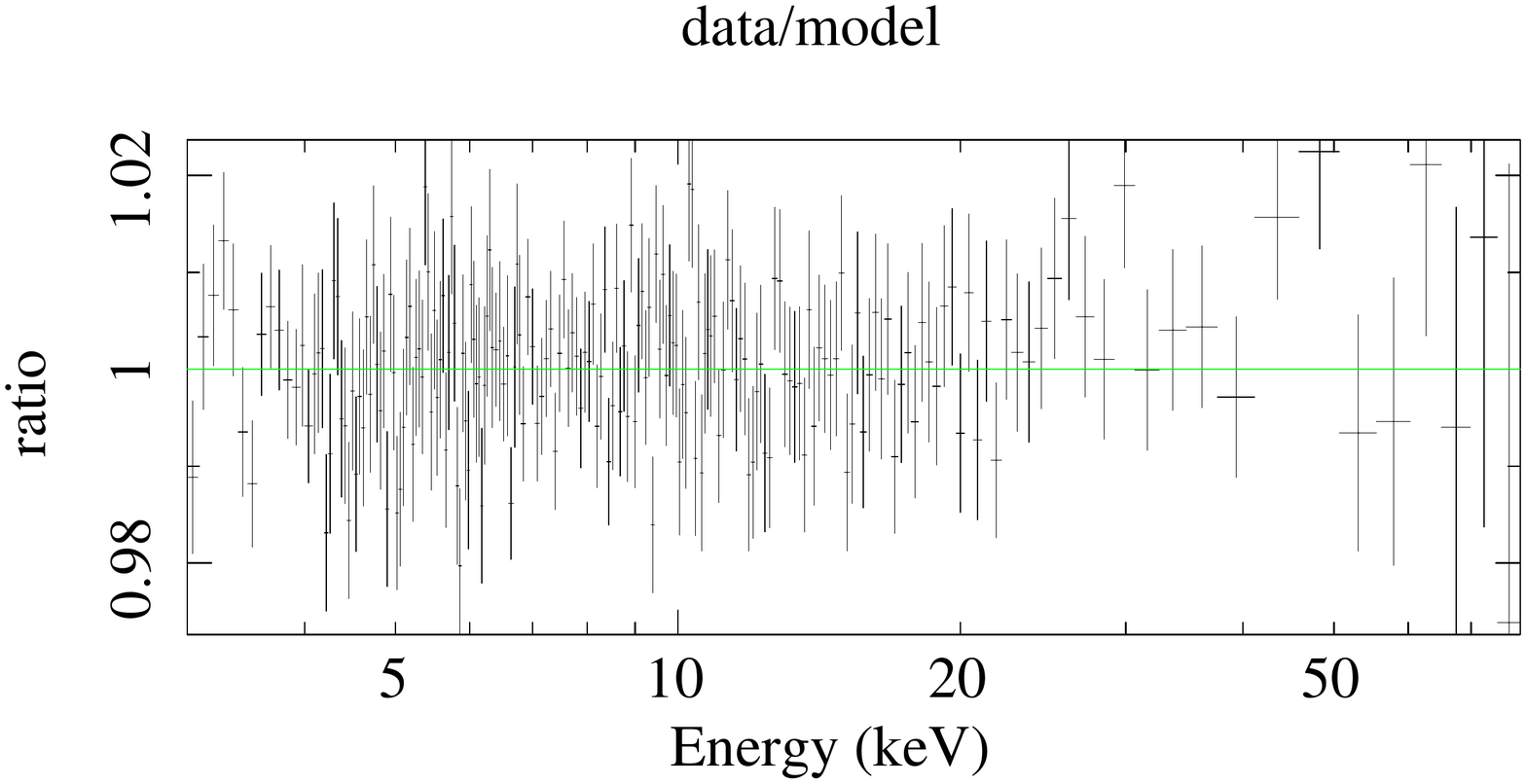}
\end{center}
\vspace{-1.4cm}
\caption{Ratio between the simulated data and the best-fit model for $\gamma = 0.99$ (left panels) and 0.75 (right panels). The spin parameter is always $a_* = 0.99$. The inclination angle is $i = 20^\circ$ (top panels), $45^\circ$ (central panels), and $70^\circ$ (bottom panels). See the text for more details. \label{f-ratio1}}
\end{figure*}

\begin{figure*}[t]
\begin{center}
\includegraphics[type=pdf,ext=.pdf,read=.pdf,width=7.9cm]{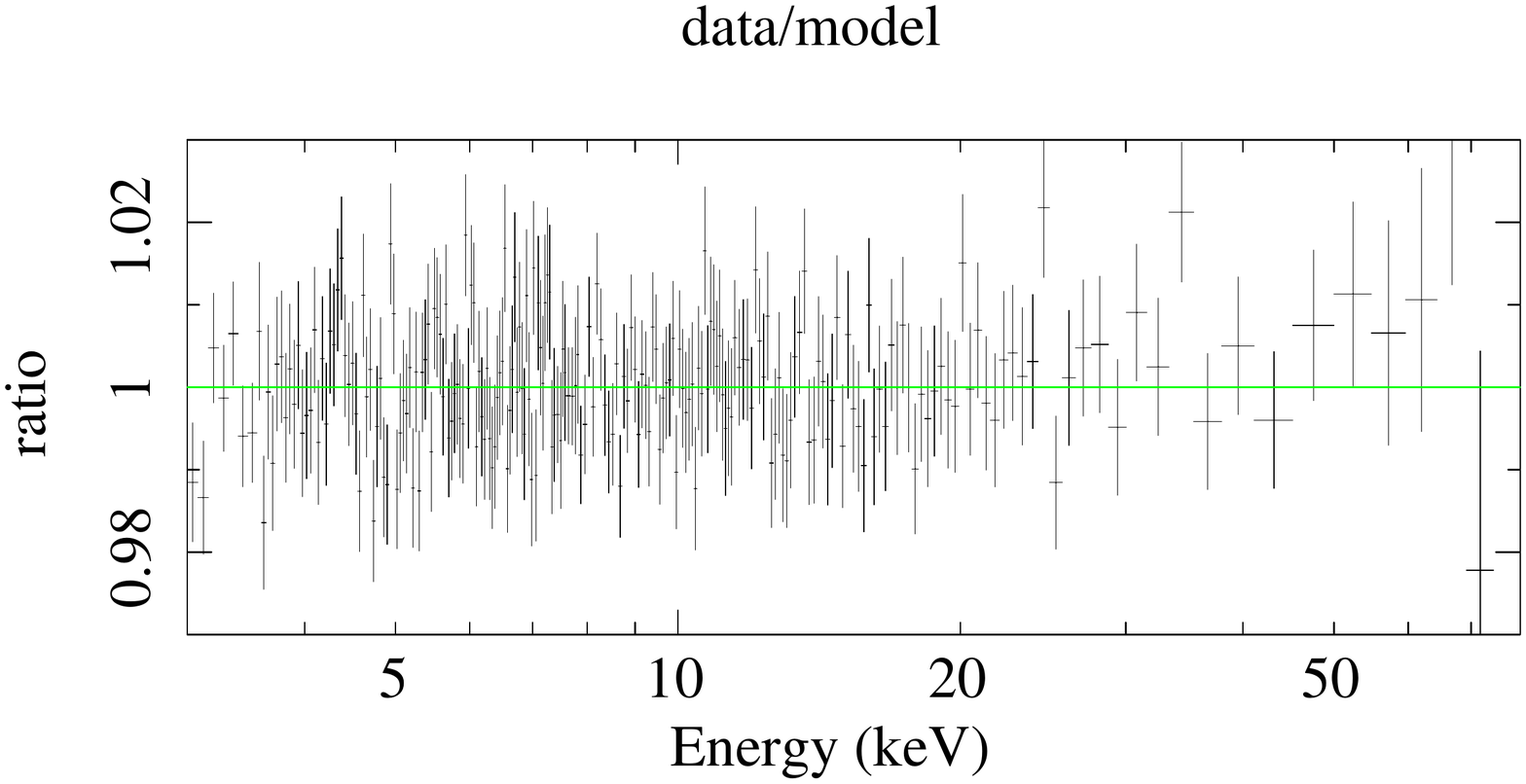}
\hspace{-0.8cm}
\includegraphics[type=pdf,ext=.pdf,read=.pdf,width=7.9cm]{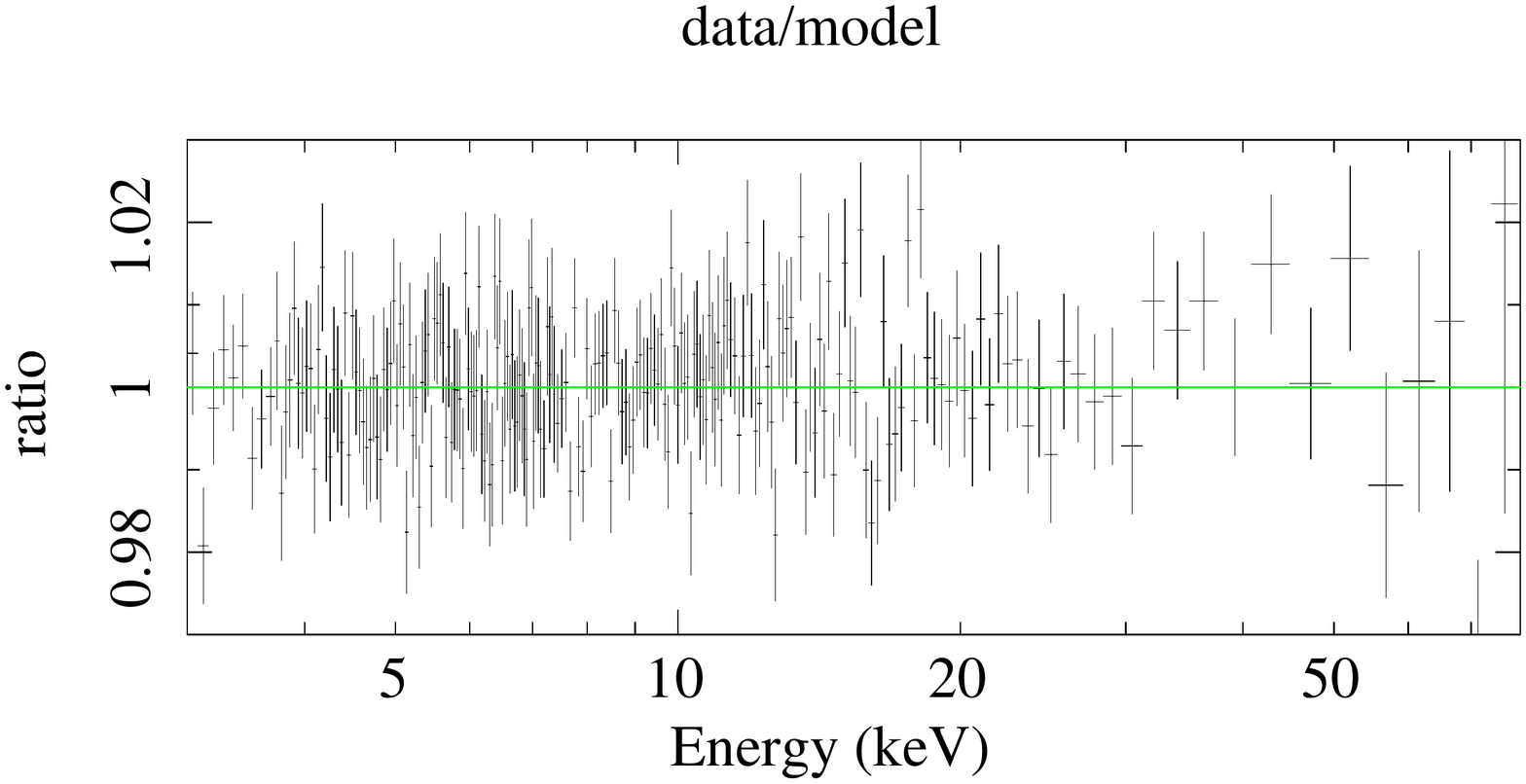}\\
\vspace{-1.9cm}
\includegraphics[type=pdf,ext=.pdf,read=.pdf,width=7.9cm]{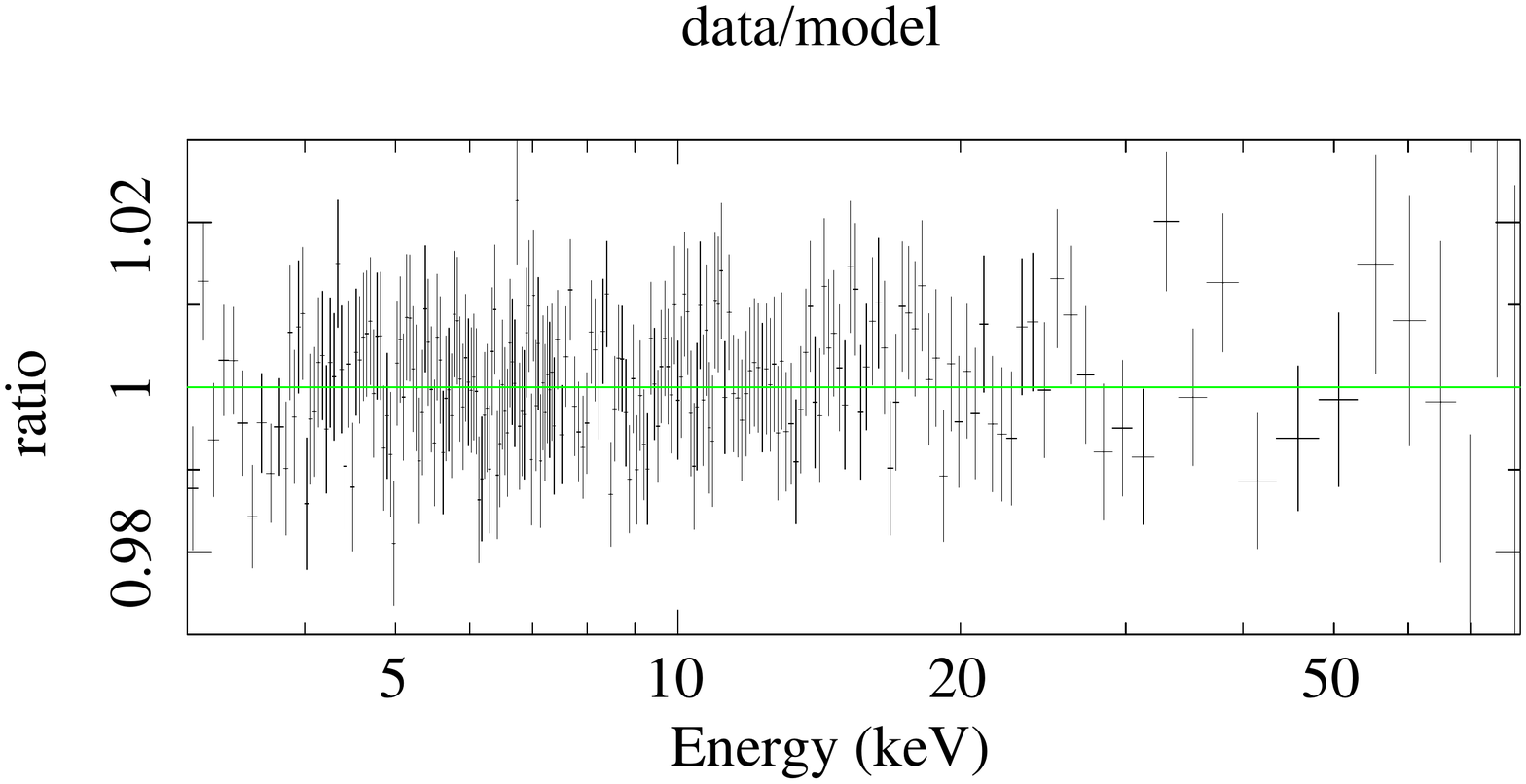}
\hspace{-0.8cm}
\includegraphics[type=pdf,ext=.pdf,read=.pdf,width=7.9cm]{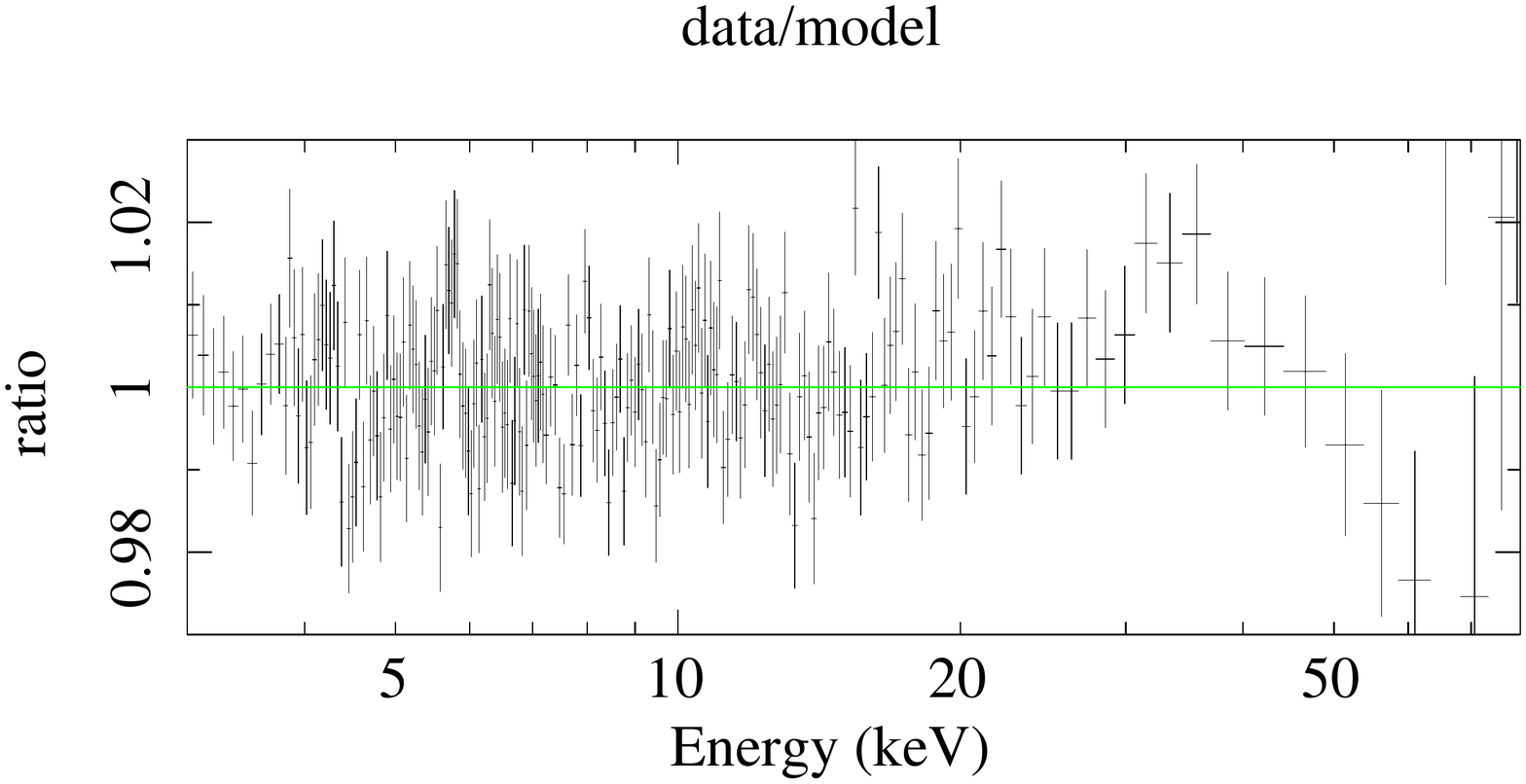}\\
\vspace{-1.9cm}
\includegraphics[type=pdf,ext=.pdf,read=.pdf,width=7.9cm]{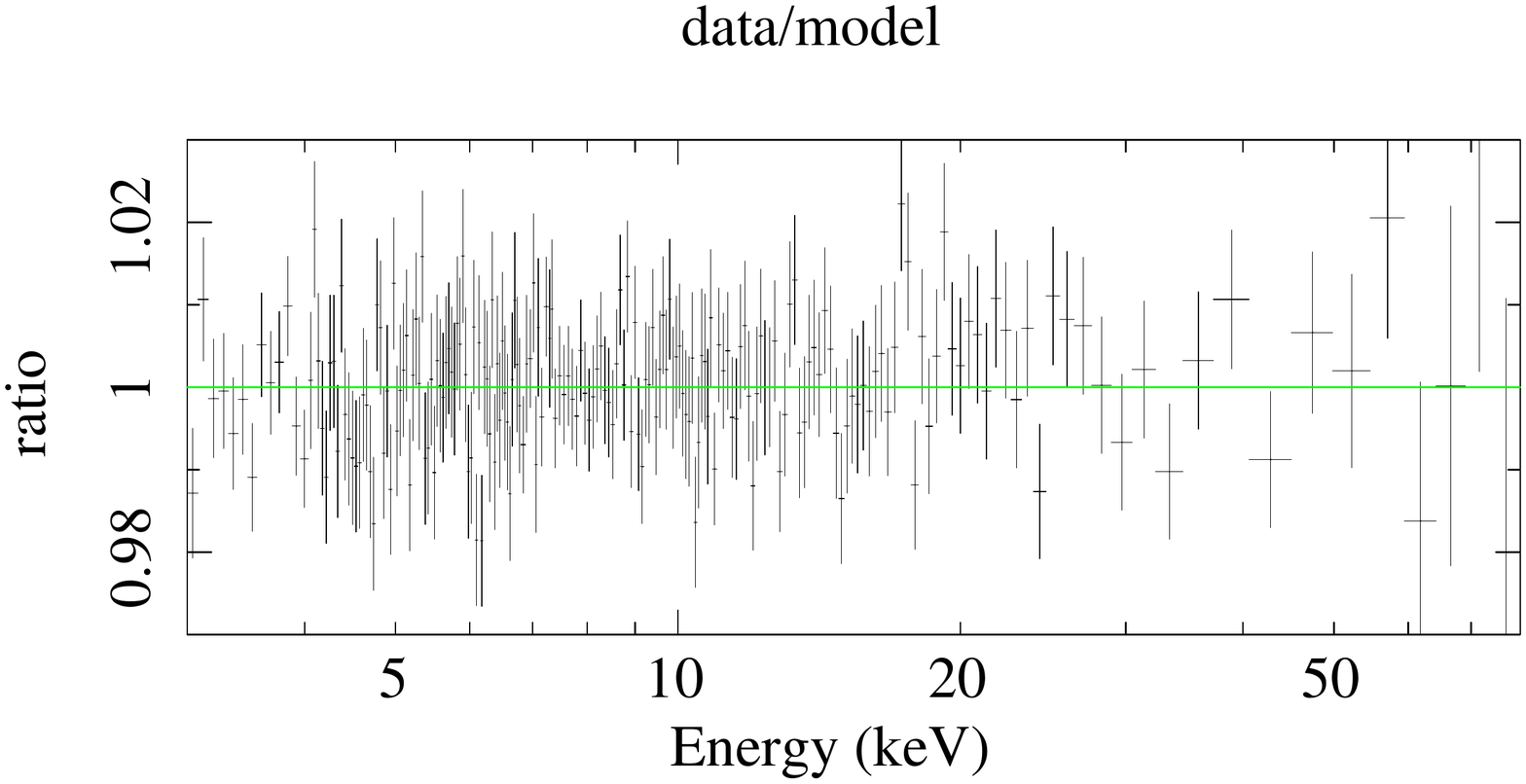}
\hspace{-0.8cm}
\includegraphics[type=pdf,ext=.pdf,read=.pdf,width=7.9cm]{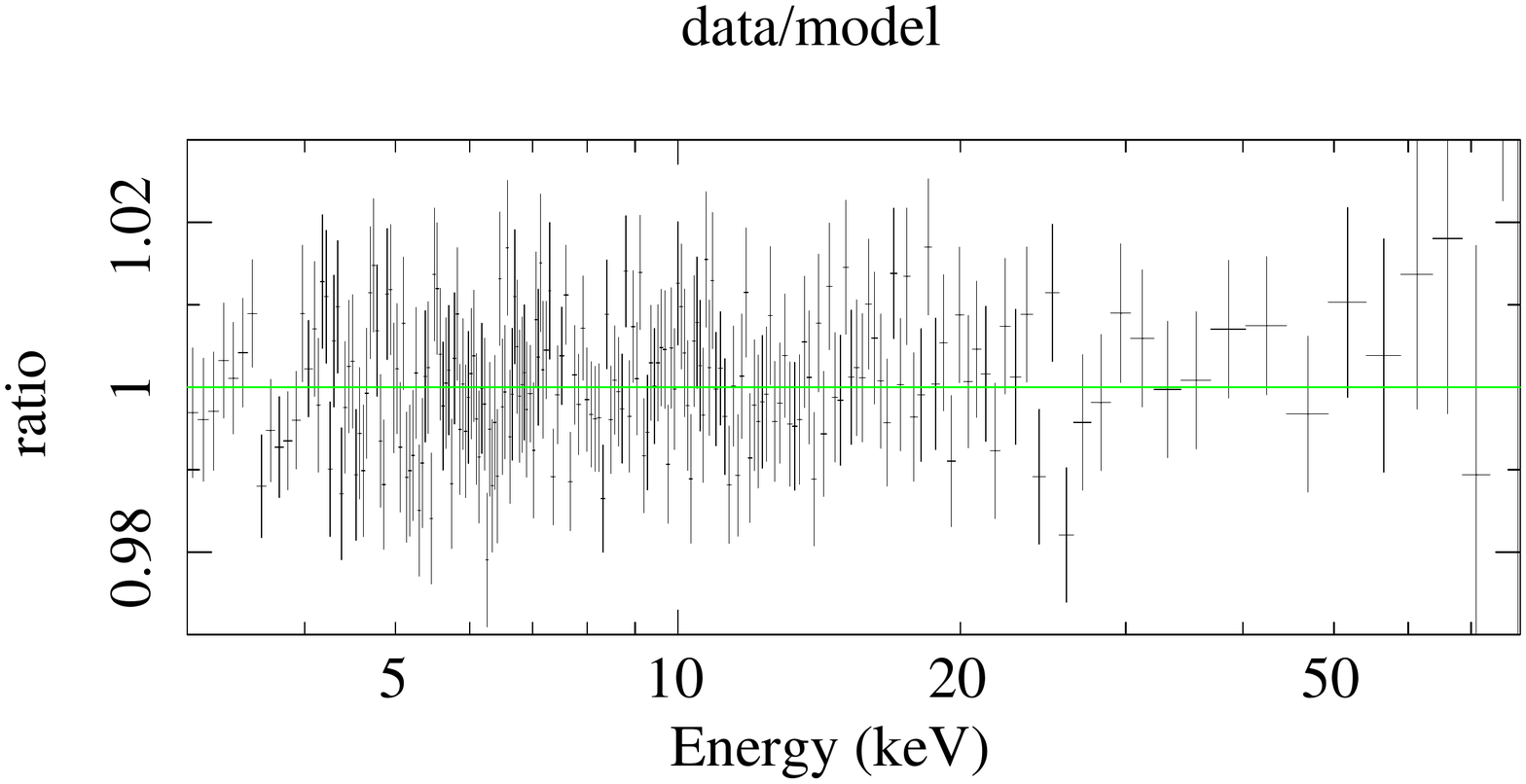}
\end{center}
\vspace{-1.4cm}
\caption{As in Fig.~\ref{f-ratio1} for $\gamma = 0.5$ (left panels) and 0.25 (right panels). The inclination angle is $i = 20^\circ$ (top panels), $45^\circ$ (central panels), and $70^\circ$ (bottom panels). See the text for more details. \label{f-ratio2}}
\end{figure*}

We have simulated twelve observations, corresponding to the case $a_* = 0.99$ for each panel shown in Fig.~\ref{f-iron} (namely $\gamma = 0.99$, 0.75, 0.5, and 0.25 and $i = 20^\circ$, $45^\circ$, and $70^\circ$). The emissivity index is always $q = 3$. Our simulated data are then fitted with a power-law component and a Kerr iron line. The latter is calculated with \verb7RELLINE7~\cite{relline}. The XSPEC model is thus \verb7POWERLAW*RELLINE7. The free parameters in the fit are $a_*$, $i$, $q$, $\Gamma$, and the normalizations of the two components. The frozen parameters are the energy line $E = 6.4$~keV and the outer edge of the disk $R_{\rm out} = 400$~$M$. The inner edge of the disk is set at the ISCO radius of the Kerr spacetime. Tabs.~\ref{t-fit1} and \ref{t-fit2} show the best-fit values of the spin parameter $a_*$, the emissivity index $q$, and of the inclination angle $i$, as well as the reduced $\chi^2$, for the twelve simulations. The tables do not show the best-fits of $\Gamma$ and the normalizations of the two components because they do not play a major rule in our results, even if a better estimate of these parameters would also help to get better measurements of $a_*$, $q$, and $i$. Figs.~\ref{f-ratio1} and \ref{f-ratio2} show the plots with the ratio between the data and the best fit model.

As we can see from Tabs.~\ref{t-fit1} and \ref{t-fit2}, the reduced $\chi^2$ of the best-fit is always close to 1, namely the fit is good. From the ratio plots, it is not evident any clear unresolved feature that could point out the difference between the JNW and Kerr spacetimes. In other words, the iron line shapes from accretion disks in JNW spacetimes can mimic pretty well those expected in Kerr spacetimes.

Note, however, that the best-fits for the input inclination angle $i = 20^\circ$ (Simulations~1, 4, 7, and 10) do not have a high spin parameter. Even for the input inclination angle $i = 45^\circ$ (Simulations~2, 5, 8, and 11), the best-fit of $a_*$ cannot be close to 1\footnote{For high inclination angles, this is not true, presumably because the stronger Doppler boosting in the JNW metric can compensate the weaker gravitation redshift.}. However, we have observations of black hole candidates with iron line shapes as expected from fast-rotating Kerr black  holes observed at low or moderate inclination angles. For example, the authors of Ref.~\cite{obs1} report the measurements $a_* = 0.97^{+0.014}_{-0.02}$ and $i = 23.7^{+6.7}_{-5.4}$~deg for the stellar-mass black hole candidate in Cygnus~X-1. A similar measurement may not be possible if Cygnus~X-1 were a JNW naked singularity. If we consider supermassive black hole candidates, the tension may be even stronger. There are indeed several sources with spin measurement very close to 1 (for example $a_* > 0.97$) and, at the same time, low or moderate viewing angle~\cite{obs2}. All these measurements seems to be difficult to fit in a JNW scenario. Of course, we can question whether a different choice of the model parameters can do the job and we can obtain JNW objects capable of mimicking fast-rotating Kerr black holes observed at a low inclination angle. The answer is no. If, for instance, we consider lower values of $\gamma$, we go to the opposite direction and it is more difficult to mimic fast-rotating Kerr black holes. Other parameters, like the emissivity index $q$, do not seem to have any impact on this issue. Note, however, that current spin measurements come with a number of caveats. See, e.g., the discussions in Ref.~\cite{rey13}. In particular, spin measurements of supermassive black holes with $a_*$ very close to 1 have to be taken with some caution. For example, current spin measurements assume that the accretion disks are infinitesimally thin and that the inner edge is at the ISCO radius. However, the mass accretion rate of several sources is probably higher than 20\% of their Eddington limit, which makes the thickness of the disk non-negligible~\cite{cr2} and permits the inner edge to be at a radius smaller than the ISCO.

Within our simple analysis based on simulated observations, we cannot arrive at any conclusive statement. Moreover, we should fit the whole reflection spectrum, not only the iron line. However, it is clear that the JNW naked singularities can unlikely well mimic fast-rotating Kerr black holes with spin parameters close to 1. The reason is that a Kerr black hole with a spin parameter very close to 1 has the ISCO radius very close to the black hole event horizon. This leads to an iron line with a tail extended to very low energies. For $a_*=1$, in Boyer-Lindquist coordinates $r_{\rm H} = r_{\rm ISCO} = M$. However, if $\gamma \neq 1$, the static limit becomes a singular surface, so the inner edge of the disk becomes $r_{\rm in} = 2\tilde{M}$ and $r_{\rm in} \rar 2M$ for $\gamma \rar 1$. There is a discontinuity in the inner edge of the disk between Kerr and JNW metrics with $a_*$ very close to 1, and this is definitively a clear observational feature to test the possibility of the existence of JNW naked singularities.

\section{Concluding remarks \label{s-con}}

Even within classical Einstein's gravity, we do not have yet any conclusive answer about the nature of the final product of gravitational collapse. Under quite general conditions, the creation of spacetime singularities is unavoidable, but this is not enough to argue whether the final outcome is a black hole or if the creation of singularities visible to distant observers are also allowed. It is usually assumed that the cosmic censorship conjecture holds, and singularities created by gravitational collapse are hidden behind an event horizon. However, no theoretical or observational confirmation is available.

In the present paper, we have presented a preliminary study to observationally distinguish black holes from naked singularities using iron line spectroscopy. We have considered the JNW metric, which is an exact solution of Einstein's gravity minimally coupled to a real massless scalar field. The metric describes the spacetime of a rotating source with a non-vanishing scalar charge $q$. The deformation parameter $\gamma$ is related to the ADM mass $M$ and the scalar charge $q$ and ranges from 1 (Kerr spacetime, vanishing scalar charge) to 0 (when $M$ is negligible with respect to the scalar charge $q$). For $\gamma < 1$, the spacetime has no horizon and there is a surface-like naked singularity where curvature invariants diverge and the spacetime is geodetically incomplete.

Our results shows that, in general, the iron line from the accretion disk of a JNW singularity can reasonably well mimic an iron line from the accretion disk around a Kerr black hole. We have simulated some observations with NuSTAR (as a prototype of a current X-ray mission) to see whether iron lines in the JNW metric can be fitted with iron lines calculated in the Kerr spacetime. The answer is positive, in the sense that the fit is good and we cannot easily distinguish the two spacetimes. However, iron line shapes in JNW spacetimes do not seem to be able to mimic iron line shapes from Kerr spacetimes with very high spins and moderate viewing angles. Since we have X-ray data that show iron line shapes as expected from fast-rotating black holes observed at moderate inclination angles, it seems that JNW spacetimes can be ruled out by those observations. A conclusive answer is beyond the scope of the present paper and would require the construction of a full reflection model (not only the iron line) and the analysis of the real data, as done in Refs.~\cite{better1,better2,better3,better4}.


\begin{acknowledgments}
This work was supported by the National Natural Science Foundation of China (Grant No.~U1531117) and Fudan University (Grant No.~IDH1512060). C.B. also acknowledges support from the Alexander von Humboldt Foundation.
\end{acknowledgments}


\appendix

\section{Calculation of iron line shapes \label{s-app}}

In Sections~\ref{s-iron} and \ref{s-sim}, the iron line shapes have been calculated with the code described in Refs.~\cite{code1,code2}. In this appendix, we briefly outline how the code works. More details can be found in the original papers.

We consider a Boyer-Lindquist coordinate system centered at the black hole and a faraway observer. The accretion disk is in the equatorial plane and the particles in the disk follow quasi-geodesics equatorial circular orbits. The 4-velocity of the particles is thus $u^\mu_{\rm e} = u^t_{\rm e} \left( 1,0,0,\Omega \right)$, where $\Omega \equiv u^\phi_{\rm e}/u^t_{\rm e}$ is the particle angular velocity (the subindex ``e'' stands for emitter). From the conservation of rest-mass $g_{\mu\nu} u^\mu_{\rm e} u^\nu_{\rm e} = - 1$, we can write
\be
u^t_{\rm e} = \frac{1}{\sqrt{- g_{tt} - 2 g_{t\phi} \Omega 
- g_{\phi\phi} \Omega^2}} \, .
\ee
$\Omega$ can be directly derived from the geodesic equations~\cite{book} 
\be
\Omega_\pm = \frac{- \partial_r g_{t\phi} 
\pm \sqrt{\left(\partial_r g_{t\phi}\right)^2 - 
\left(\partial_r g_{tt}\right) \left(\partial_r g_{\phi\phi}\right)}}{\partial_r g_{\phi\phi}} \, , 
\ee
where the upper (lower) sign refers to corotating (counterrotating) orbits, namely orbits with angular momentum parallel (antiparallel) to the spin of the central object. In the case of the JNW metric, the expression for $\Omega$ is
\be
\Omega_\pm &=& A
\frac{M \Sigma' A \omega + \sqrt{M \Delta' \sin^2\theta \left( \Sigma \Sigma' \left( r - \tilde{M} \right) 
- M \Sigma'^2 \right) }}{M \Sigma' A^2 \omega^2 -
\Delta' \sin^2\theta \left[ \Sigma \left( r - \tilde{M} \right) 
- \Sigma' M\right]} \, ,
\ee
where we have defined $\Sigma' = r^2 - a^2\cos^2\theta$, $\Delta' = r^2 - 2 \tilde{M} r + a^2\cos^2\theta$, and $A = ( 1 - 2 \tilde{M} r / \Sigma )^\gamma$. The 4-velocity of the distant observer is $u^\mu_{\rm obs} = \left( 1,0,0,0 \right)$.

The observer has an image plane with a grid, and at every point of the grid we have a photon perpendicular to the image plane. The iron line spectrum is obtained by calculating the image of the accretion disk and integrating over the image plane. The image of the accretion disk is calculated by firing photons from the image plane backwards in time to the accretion disk. We write the photon initial conditions (initial position and initial 4-momentum) in the coordinate system centered at the black hole. For every photon in the grid, we calculate backwards in time the photon trajectory from the detection point in the image plane to the emission point in the accretion disk in the equatorial plane. Photons that miss the disk (they hit the back hole or cross the equatorial plane between the black hole and the inner edge of the accretion disk) do not contribute to the iron line spectrum (here we only consider the primary image of the disk).

The code computes the photon flux number density as measured by a distant observer
\be
N (E_{\rm obs}) &=& \frac{1}{E_{\rm obs}} 
\int I_{\rm obs} \left( E_{\rm obs} \right) d \omega 
= \frac{1}{E_{\rm obs}} 
\int g^3 I_{\rm e} \left( E_{\rm e} \right) d \omega \, ,
\ee
where $E_{\rm obs}$ and $I_{\rm obs}$ are, respectively, the photon energy and the specific intensity of the radiation measured by the distant observer, $E_{\rm e}$ and $I_{\rm e}$ are the same quantities in the rest-frame of the particles in the accretion disk, and $d\omega$ is the line element of the solid angle subtended by the image of the disk on the observer's sky. $g$ is the redshift factor 
\be
g = \frac{E_{\rm obs}}{E_{\rm e}} = \frac{k_\mu u^\mu_{\rm obs}}{k_\nu u^\nu_{\rm e}} \, ,
\ee
where $k^\mu$ is the photon 4-momentum. If we plug the expressions of $u^\mu_{\rm obs}$ and $u^\mu_{\rm e}$, we find 
\be
g = \frac{\sqrt{- g_{tt} - 2 g_{t\phi} \Omega 
- g_{\phi\phi} \Omega^2}}{1 + \lambda \Omega} \, 
\ee
where $\lambda = k_\phi/k_t$ is a constant of motion along the photon path. $I_{\rm obs}/E_{\rm obs}^3 = I_{\rm e}/E_{\rm e}^3$ follows from Liouville's theorem.

For every photon in the grid of the image plane of the distant observer, we find the photon emission point in the disk and we calculate the redshift factor $g$ and the emission radius $r_{\rm e}$. The disk's emission is assumed to be monochromatic and isotropic with a power-law radial profile
\be
I_{\rm e} \left( E_{\rm e} \right) \propto 
\frac{\delta \left( E_{\rm e} - E_{\rm K\alpha} \right)}{r_{\rm e}^q} \, ,
\ee
where $E_{\rm K\alpha} = 6.4$~keV is the rest-frame energy of the emission line and $q$ is the emissivity index.

The ray-tracing calculations provide a photon count with a certain photon energy for every point of the grid of the image plane. After integrating over the whole disk's image, we obtain the iron line shapes reported in Sections~\ref{s-iron} and \ref{s-sim}.


\end{document}